# Volatile Element Chemistry during Metamorphism of Ordinary Chondritic Material and Some of its Implications for the Composition of Asteroids

By

Laura Schaefer

And

Bruce Fegley, Jr.

Planetary Chemistry Laboratory
Department of Earth and Planetary Sciences
Washington University
St. Louis, MO 63130-4899
laura_s@wustl.edu
bfegley@wustl.edu

**Proposed Running Head:** Volatile element metamorphic chemistry on asteroids

**Corresponding Author:**

Laura Schaefer

Campus Box 1169

Department of Earth and Planetary Science

Washington University

One Brookings Dr.

St. Louis, MO 63130-4899

laura_s@wustl.edu

Phone: 314-935-6310

Fax: 314-935-7361

**Abstract:**

We used chemical equilibrium calculations to model thermal metamorphism of ordinary chondritic material as a function of temperature, pressure, and trace element abundance and use our results to discuss volatile mobilization during thermal metamorphism of ordinary chondrite parent bodies. The calculations include ~ 1,700 solids and gases of 40 elements. We compiled trace element abundances in H, L, and LL chondrites for the elements Ag, As, Au, Bi, Cd, Cs, Cu, Ga, Ge, In, Pb, Rb, Sb, Se, Sn, Te, Tl, and Zn, and identified abundance trends as a function of petrographic type within each class. We found that abundance patterns within the H- and L-chondrites are consistent with mobilization of volatile elements in an onionshell-type parent body. LL-chondrites have more complex abundance patterns that may support a rubble-pile model for the LL-chondrite parent body. We calculated volatility sequences for the trace elements in the ordinary chondritic material, which differ significantly from the solar nebula volatility sequence.

## 1. Introduction

Recently we used chemical equilibrium calculations to model thermal outgassing of ordinary chondritic material as a function of temperature, pressure and bulk composition (Schaefer and Fegley, 2007a). This work considered the chemistry of 20 major, trace, and volatile elements (Al, C, Ca, Cl, Co, Cr, F, Fe, H, K, Mg, Mn, N, Na, Ni, O, P, S, Si, and Ti). We discussed the results relevant to outgassing and atmospheric formation in our earlier paper. However, the chemical equilibrium calculations in Schaefer and Fegley (2007a) also provide the foundation for modeling chemistry of the volatile major and trace elements during thermal metamorphism of ordinary chondritic material. The reason for this is as follows.

During thermal metamorphism, volatile elements may vaporize into the gas phase and be transported in the parent body. The details of this process depend to some extent on the molecular speciation of the elements in the gas phase. In turn, this speciation depends on the fugacities (equal to partial pressures for ideal gases) of more abundant elements such as oxygen, sulfur, hydrogen, nitrogen, chlorine, and fluorine in the gas phase. The chemistry of volatile trace elements (e.g., Pb, Bi, In, Tl, etc.) depends upon that of the major elements, but does not alter major element chemistry because the volatile trace elements are much less abundant than the major elements. However, the chemistry of volatile trace elements with similar abundances may be interconnected.

Our work addresses three important questions. First, do elemental volatilities during ordinary chondrite metamorphism differ from those during condensation in the solar nebula? This question is important because cosmochemists presently use the nebular volatility sequence to discuss metamorphic chemistry of volatile elements.

Second, does thermal metamorphism affect the abundances and chemistry of volatile elements in ordinary chondrites? Third, do volatile element abundances vary inside thermally metamorphosed asteroids?

We focus on carbon and volatile trace elements (Ag, As, Au, Bi, Cd, Cs, Cu, Ga, Ge, In, Pb, Rb, Sb, Se, Sn, Te, Tl, Zn) in this paper for two reasons. First, thermal metamorphism was isochemical with respect to the major rock-forming elements such as Ca, Mg, Al, Si, Fe, and Ti (McSween et al., 1988). Second, Schaefer and Fegley (2007a) found that the abundances of F, Cl, and S, which are also volatile elements, are unaltered by thermal metamorphism. These elements are mainly in minerals throughout the range of metamorphic temperatures instead of being in the gas phase.

Our paper is organized as follows. Section 2 gives background information on ordinary chondrites and thermal metamorphism. Section 3 discusses the trace element abundances in H-, L-, and LL-chondrites and abundance trends as a function of petrologic type. Section 4 describes our thermodynamic calculations. In Section 5, we discuss the results of our thermodynamic calculations for the trace elements. We take our discussion of carbon from Schaefer and Fegley (2007a), who discuss results for 20 major and minor elements. In section 6, we describe the calculated volatility sequence and compare it to solar nebular volatility and trace element mobility in ordinary chondrites. We also discuss, using our results, possible mechanisms for producing the observed trends in the trace element abundances. Section 7 gives our answers to the questions posed in the introduction, summarizes our research, and suggests how it can be extended. Schaefer and Fegley (2007b) presented preliminary results of this work.

## 2. Some Background Information about Ordinary Chondrites

Ordinary chondrites make up about 97% of all chondrites and are the most abundant type of meteorites. They are unmelted stony meteorites that contain metal, silicate, and sulfide minerals in varying proportions. The ordinary chondrites are divided into three groups (H, L, LL) on the basis of the total elemental abundance of iron and the abundance of iron metal. The H-chondrites (~49% by number of all ordinary chondrites) have high total iron and high iron metal. The L-chondrites (~36%) have low total iron, and the LL-chondrites (~14%) have low total iron and low iron metal. To first approximation, the ordinary chondrites are composed of a few abundant silicate minerals, iron metal, and troilite (FeS). For example, the average mineralogical composition (in mass percent) of H-chondrites is 35.0% olivine $(Mg,Fe)_2SiO_4$, 26.2% pyroxene $(Mg,Fe)SiO_3$, 18.0% Fe-rich metal, 9.6% feldspar ($CaAl_2Si_2O_8$ – $NaAlSi_3O_8$ – $KAlSi_3O_8$), 5.5% FeS, 4.1% diopside $CaMgSi_2O_6$, and 1.6% other minerals. These figures are rounded mass percentages from McSween et al. (1991). The mineralogical compositions of the L and LL chondrites are similar to that of the H-chondrites (see Table 3 of Schaefer and Fegley, 2007a or McSween et al. 1991).

As mentioned in our prior paper (Schaefer and Fegley, 2007a), chondrites are samples of material from the solar nebula. Different types of chondritic material (e.g., carbonaceous, ordinary, enstatite) analogous to different types of chondritic meteorites are generally believed to be the building blocks of the Earth and other rocky bodies. Chemical equilibrium calculations, cosmochemical models, and geochemical data indicate that ordinary chondritic material was abundant in the inner solar nebula where the terrestrial planets formed, and that the Earth is broadly chondritic in composition (e.g., see Barshay, 1981; Hart and Zindler, 1986; Kargel and Lewis, 1993; Larimer, 1971; Lewis and Prinn, 1984; Lodders, 2000; Lodders and Fegley, 1997; Wänke, 1981).

Although they are samples of nebular material, the ordinary chondrites are not unaltered, and underwent dry thermal metamorphism at 400˚ – 1,000˚ C at relatively low pressures (< 2 kilobars) on their parent bodies (McSween et al., 1988). A numerical scale running from 3 to 6 for the H-, L-, and LL-ordinary chondrites denotes the degree of metamorphic alteration. For example, the type 3 (unequilibrated) chondrites with metamorphic temperatures ranging from 400˚ – 600˚ C are the least metamorphically altered while the type 4 – 6 (equilibrated) chondrites with metamorphic temperatures ranging from 600˚ – 1000˚ C are more altered. (The terms equilibrated and unequilibrated refer to the homogeneity of ferromagnesian silicates, but do not imply thermochemical equilibrium or lack thereof.)

The observed variations among unequilibrated chondrites are sufficiently great that they are further sub-divided into types 3.1 – 3.9. The large variations are not surprising because chemical equilibria and chemical reaction rates vary exponentially with temperature, i.e., log $K_{eq}$ or log $k_{rate}$ vs. 1/T (in Kelvins) is linear where $K_{eq}$ is the equilibrium constant and $k_{rate}$ is the rate constant for a chemical reaction. The inverse temperature range ($10^4/T = 11.45 – 14.86$) across the type 3 chondrites is much greater than across any other petrologic type and is almost as wide as the inverse temperature range across all equilibrated chondrites ($10^4/T = 7.85 – 11.45$). On this basis, greater variations are expected within the type 3 chondrites than within any other single petrologic type.

The effects of thermal metamorphism on the chemistry of ordinary chondrites have been studied for over 40 years. As noted earlier, these studies show that thermal metamorphism was isochemical with respect to major rock-forming elements. However, thermal metamorphism apparently altered the abundances, distribution with petrologic type, and speciation of *some* volatile trace elements in ordinary chondrites. For example, Otting and Zähringer (1967) showed that the abundances of the noble gases ($^{36}Ar$, $^{132}Xe$) varied inversely with petrologic type. Carbon contents also vary inversely with petrologic type from type 3.1 – 3.9, although the variation is not as clear from types 4 – 6. Tandon and Wasson (1968) showed that the In concentration varied inversely with petrologic type in L-chondrites. On the other hand Wolf and Lipschutz (1998) found no statistically significant abundance trends for several volatile trace elements (Co, Rb, Ag, Se, Te, Zn, Cd, Bi, In) with petrologic type in H4 – H6 chondrites.

If thermal metamorphism is responsible for elemental abundance variations as a function of petrologic type, it has important implications for elemental abundances on asteroids, which are the meteorite parent bodies. The onionskin model for meteorite metamorphism on asteroids postulates that the more metamorphosed type 6 – 4 equilibrated chondrites come from the hotter, higher pressure interior of an asteroid and that the less metamorphosed type 3.9 – 3.1 unequilibrated chondrites come from the colder, lower pressure surface regions. Dreibus and Wänke (1980) proposed that volatile elements were transported from the hotter interiors to the cooler outer layers of meteorite parent bodies. Other workers have made similar suggestions. If such ideas are correct, our chemical equilibrium calculations should predict the radial variation of volatile element concentrations and host phases inside asteroidal parent bodies. These predictions are testable using geochemical analyses conducted by remote sensing and in situ methods. We return to this idea in Section 6 where we compare our calculations with observations.

**3. Trace Element Abundances in Ordinary Chondrites**

We compiled trace element abundances for 20 volatile trace elements (Rb, Cs, Cu, Ag, Au, Zn, Cd, Ga, In, Tl, Ge, Sn, Pb, As, Sb, Bi, Se, Te, Br, and I) in H-, L-, and LL- chondrite falls. We used the METBASE meteorite database (Koblitz, 2005) and additional references to do this (Binz et al., 1976; Dennison and Lipschutz, 1987; Friedrich et al., 2003, 2004; Huston and Lipschutz, 1984; Keays et al., 1971; Lingner et al., 1987; Luck et al., 2005; Neal et al., 1981; Walsh and Lipschutz, 1982; Wang et al., 1999; Wolf and Lipschutz, 1995; Wolf et al., 1997). Falls are meteorites that were recovered at the time they were observed to fall. Conversely, finds are meteorites that were recovered some time after they fell to Earth. The overall range, means, and medians for the trace elements in H-, L-, and LL-chondrite falls of all petrologic types are given in Tables 1-3. Elements in bold typeface have means which are more than 50% larger than their respective medians. We discuss this further below.

Variations of trace element abundances between petrologic types are discussed in Section 3.1 and are shown in Figures 1 – 7. We examined the available analyses to determine if the trends previously observed in trace element abundances with petrologic type (Morgan et al., 1985; Sears and Weeks, 1986; Lingner et al., 1987; Keays et al., 1971) are real or due to limited or biased sample bases (Kallemeyn et al., 1989; Wolf and Lipschutz, 1998). We used only data for chondrite falls because many minor and trace elements are susceptible to leaching and contamination in finds (Mason, 1971). However, not all falls are immune from these problems and not all finds are affected by them. For example, some specimens of the Beardsley H5 chondrite were collected a year or so after it fell and are depleted in alkalis due to weathering. In contrast, the Potter L6 chondrite is a find with a terrestrial residence time longer than 20,000 years but it is not depleted in K relative to other L-chondrites (Goles, 1971).

Following Wolf and Lipschutz (1998), we removed analyses for a few H chondrite falls from consideration: Rose City (H5), Yanzhuang (H6), Cangas de Onis (H5), Fayetteville (H4), Leighton (H5), Pantar (H5), Tynes Island (H4), and Weston (H4). Post-accretional shock heating strongly modified Rose City and Yanzhuang. Also, Rose City is a brecciated and veined meteorite (see Figure 5 in Mason and Wiik, 1966) and a representative sample would consume a large amount of the meteorite. For example, the reported oxygen abundance of 27.7 wt. % is too low for H-chondrites (Mason and Wiik, 1966; Schaefer and Fegley, 2007a). The remaining four meteorites are gas-rich regolith breccias, which are enriched in volatile elements.

As we discuss below, trends were observed within H3 chondrites for Bi, In, Tl (and possibly Ag and Cd). The concentrations of these elements decreased significantly as petrologic type increased from 3.4 to 3.9. However, these trends are based on multiple analyses for only three individual meteorites: Sharps (H3.4), Tieschitz (H/L3.6), and Bremervörde (H/L3.9). The H/L-chondrites, being intermediate between H- and L-chondrites, are classified as such on the basis that they have lower abundances of siderophile elements than true H-chondrites (Kallemeyn et al., 1989). It is this that causes the apparent drop in abundances for Rb, Cs, Au, Zn, Ga, Ge, and also As in H3 chondrites as compared to the higher petrologic types. Whether a definitive trend for the H3 sub-type can be established based solely on data from these 3 meteorites is debatable.

We included data for all shock types; however, it has been shown that there are compositional trends with shock type for the L-chondrites (Neal et al., 1981; Walsh and Lipschutz, 1982; Huston and Lipschutz, 1984), with abundances significantly lower in more highly shocked (S4-S6) chondrites, than in mildly shocked (S1-S3) chondrites. A similar trend is not observed for H-chondrites (Lingner et al., 1987). The analyses which we used are split approximately evenly between lightly and heavily shocked meteorites.

We calculated mean and median abundances for each trace element in each petrologic type and for each class (H, L, LL) as a whole. We found that trends in mean abundances with petrologic type, where observed, are typically subtle and can be easily obscured by inclusion of one or two anomalously large or small analyses. We believe that median abundances track trends much better as they are less susceptible than means to the effects of such anomalous analyses. For instance, the increase of Tl abundances with increasing petrologic type in H4 – H6 chondrites, also observed by Wolf and Lipschutz (1998), is only apparent for the mean abundances; the median abundances are constant throughout the H4 – H6 chondrites, which seems easier to understand.

In Section 3.1, we discuss the observed trends, or lack there of, for trace element abundances as a function of petrologic type in the ordinary chondrites. We organize the discussion according to periodic table groups. In Section 3.2, we compare the observed trends within each ordinary chondrite class and across the classes to determine if there are any correlations.

### *3.1 Trace Element Abundances Across Petrologic Type*

*3.1.1. Group IA: Rb, Cs.* Schaefer and Fegley (2007a) discussed Na and K abundances in ordinary chondrites. Lithium is not a volatile trace element and we do not consider it.

That leaves rubidium and cesium; their abundances in H-, L-, and LL-chondrite falls are shown in Fig. 1. Curtis and Schmitt (1979) analyzed Rb in mineral separates of three L6 chondrites [Alfianello, (Colby, Wisconsin) and Leedey]. They found that the feldspar contains 48 – 65% of the total Rb in these chondrites. However, very little Rb was found in olivine, pyroxenes, troilite, or chromite. Gast (1962) studied leaching of K, Rb, and Cs from the Beardsley H5 chondrite and found that 9.5% of all K, 27% of total Rb, and 75% of total Cs were dissolved. He suggested that the alkalis are located in two different sites – feldspar (containing most of all K and Rb), and an unidentified water-soluble mineral (containing most of all Cs, and some Rb). Smales et al. (1964) studied aqueous leaching of Rb and Cs from the Bremervörde H/L3.9, Crumlin L5, Homestead L5, Holbrook L6, and Modoc L6 chondrites (all falls). They found that leaching removed more total Cs than Rb, e.g., 19.2% of Cs and only 9.7% of Rb from the Crumlin L5 chondrite. Wieler et al. (2000) analyzed halite (NaCl) and sylvite (KCl) in the Monahans (1998) H5 chondrite and found ~ 1000 μg/g Rb in the sylvite where it substitutes for potassium. A significant amount of total Rb in Monahans (1998) is probably in the sylvite. There do not seem to be any abundance trends with petrologic type for Rb in H- or L-chondrites because the ranges for all petrologic types overlap, and there is no systematic trend in the means or medians with petrologic type. Although ranges overlap, the unequilibrated LL3 chondrites have a significantly higher mean and median than the equilibrated LL4 – 6 chondrites. Additionally, a slight trend is seen in mean and median Rb abundances with petrologic type ($4 < 5 < 6$).

Curtis and Schmitt (1979) found that feldspar contains 19 – 35% of the total Cs in the three L-chondrites they analyzed. However, they also found significant amounts of Cs in olivine, pyroxenes, troilite, and chromite. All of the Cs in the Leedey L6 chondrite is in the minerals they analyzed, but 65% of all Cs in the Colby, Wisconsin chondrite and 31% of all Cs in the Alfianello chondrite are not in the minerals analyzed. Mason and Graham (1970) found ~ 400 ng/g Cs in feldspar in the Modoc L6 chondrite. Feldspar has a normative abundance of 10.8% in Modoc (Mason and Wiik, 1967), and accounts for about 43 ng/g Cs in the whole rock. The mean and median Cs contents in Modoc are 76 ± 14 ng/g and 80 ng/g, respectively, calculated from the data of Gast (1960), Keays et al. (1971), and Smales et al. (1958, 1964). Thus, feldspar contains about 55% of all Cs in Modoc. These results for L6 chondrites and the leaching experiments of Gast (1962) and Smales et al. (1964) on H- and L-chondrites indicate that the Cs in ordinary chondrites is probably distributed between water-soluble and insoluble phases with feldspar being the likely candidate for the insoluble phase. However, $Cs^+$ is larger than $Rb^+$ so less of the total Cs is probably found in feldspar than is the case for Rb.

As shown in Figure 1, the average Cs abundances apparently decrease with increasing petrologic type in H- and L-chondrites (4 > 5 > 6). Wolf and Lipschutz (1998) also concluded that the Cs abundance decreased with increasing petrologic type in equilibrated (4 – 6) chondrites. The L3 chondrites fall into this trend, but it is unclear if the H3 chondrites fit into the same pattern. The LL-chondrites also show decreasing Cs abundance with increasing petrologic type for types 3 – 5; however, type LL6s have a higher mean Cs abundance than the other petrologic types (LL6 > LL3 > LL4 > LL5).

*3.1.2. Group IB: Cu, Ag, Au.* Abundances of Cu, Ag, and Au in H-, L-, and LL-chondrite falls are shown in Fig. 2. Copper abundances are essentially constant across petrologic types for H-, L-, and LL-chondrites. A high value of 239 μg/g for Cu in the Benton LL6 chondrite is excluded from the range, mean, and median because Schmitt et al. (1972), who measured this value, also discarded it from their calculations. Excluding this value, the range for Cu in LL6 chondrites is comparable to that for LL3 chondrites.

Silver is present in metallic and sulfide phases in meteorites (Mason, 1971). Silver abundances in ordinary chondrites are scattered (see Fig. 2). The Ag abundances in carbonaceous chondrites also are scattered (Palme et al., 1988). In H-chondrites, no immediate pattern is obvious for the mean Ag contents. However, the large ranges in the H5 and H6 chondrites are due to two large analyses for Gnadenfrei H5 (578 and 1870 ng/g) and one analysis for Naoki H6 (511 ng/g). If these analyses are rejected, the mean abundances drop to be much closer to the median abundances. Then the pattern H3 > H4 – H6 is apparent. A similar pattern is visible for the L- and LL-chondrites. Additionally, within the H3 subgroup, there is a possible correlation of Ag abundance with petrologic type from 3.4 to 3.9.

Gold is highly siderophile and is found primarily in the metallic phase, but there are significant amounts of Au in chromite, feldspar and phosphate minerals in ordinary chondrites (Rambaldi et al., 1978). There appears to be a slight increase of Au abundances with decreasing petrologic type for all three ordinary chondrite groups, but the trend is relatively insignificant. It is the most pronounced for H4-H6 chondrites, but Au abundances in H3 chondrites are significantly lower. However, this is likely due to the lower siderophile abundances in the H/L-chondrites. An average of three values for H3 chondrites gives 230 ng/g, similar to the higher petrologic types. Gold abundances increase slightly from LL- to L- to H-chondrites, in accord with the increase in Fe metal.

*3.1.3. Group IIB: Zn, Cd, Hg*. Abundances of Zn and Cd in H-, L-, and LL-chondrite falls are shown in Fig. 3. In ordinary chondrites, zinc is found distributed throughout silicate phases and chromite; very little is found in troilite (Curtis and Schmitt, 1979). Cadmium seems to be concentrated in troilite, but also has similar geochemical behavior to zinc, so some may also occur in silicates (Mason, 1971). Zinc abundances do not show a significant trend with petrologic type. The Zn concentration in H3 chondrites is lower than for the higher petrologic types, possibly due to the inclusion of type H/L3 chondrites. However, consideration of only H3 chondrites does not raise the mean value at all. The mean Zn concentration in L4 chondrites is significantly higher than for the other petrologic types, due to a single large analysis for Rio Negro (291 μg/g); however, the median abundance appears to follow the trend. Zinc abundances in LL-chondrites appear relatively constant.

Cadmium abundances show significant scatter as seen in the very large abundance ranges. Cadmium abundances are significantly enriched in unequilibrated ordinary chondrites. In H-chondrites, the mean and median Cd abundances differ significantly for H4 and H6 chondrites, due to a single large analysis for both types. If we neglect the analyses for the Forest Vale H4 (1240 ng/g) and Kernouve H6 (1040 ng/g) chondrites, the Cd abundance ranges and means in H4 and H6 chondrites are comparable to those of H5 chondrites. Then the H3 chondrites would have significantly higher means and medians than the higher petrologic types. We have only three analyses for H3 chondrites, so it is not really possible to determine if there is a trend of Cd abundance within the H3 subtype. The large Cd abundance range in L6 chondrites is primarily due to a single large value for the Nejo L6 chondrite (875 ng/g). If we reject this analysis, the mean Cd abundance in L6 chondrites is ~50 ng/g, much closer to the median. Then, the L3 and L4 chondrites have higher Cd abundances than the L5 and L6 chondrites. It is interesting that for the H- and L-chondrites, the lower petrologic types have much tighter abundance ranges than the higher petrologic types. In LL-chondrites, however, the mean Cd abundances increase with decreasing petrologic type (3>4>5>6).

Mercury is not included in our abundance compilation or thermodynamic calculations because its meteoritic abundance is too uncertain (Lodders, 2003). Mercury is extremely volatile and terrestrial contamination of meteorites is a major problem (Lipschutz and Woolum, 1988). For example, Stakheev et al. (1975) showed that the Hg abundance of meteorites increases with storage time due to sorption.

*3.1.4. Group IIIA: Ga, In, Tl.* Neither boron nor aluminum is a volatile trace element and we do not consider them here. However, Schaefer and Fegley (2007a) discussed Al abundances in ordinary chondrites. The abundances of Ga, In, and Tl in H-, L-, and LL-chondrite falls are shown in Fig. 4. Gallium occurs in metal, oxide, and sulfide in ordinary chondrites, e.g., see Allen and Mason (1973) and Mason (1971). The abundance of Ga in metal increases with petrologic type in L- and LL-chondrites, although the same trend is not observed in H-chondrites (Chou and Cohen, 1973; Chou et al., 1973). Gallium abundances are fairly constant across petrologic type for H- and L-chondrites; although there is a slight increase in Ga abundances from H6 to H4 chondrites. The Ga abundances in LL-chondrites decrease in the sequence LL4 < LL5 < LL6, but LL3 chondrites have Ga abundances similar to LL6 chondrites.

Indium occurs in the non-magnetic fraction (silicates + troilite) in meteorites and it is enriched in the dark (noble gas rich) portion versus the light (noble gas poor) portion of brecciated chondrites (Mason, 1971). Indium abundances for H4, H5 and H6 chondrites are approximately constant but those in H3 chondrites are significantly higher. The In abundances in H3 chondrites also decrease with increasing petrologic type from H3.4 to H3.9. The large range of In abundances in H5 chondrites is due to two large analyses, for the Castalia (56 ng/g) and Searsmont (22.8 ng/g) meteorites. If we neglect these two analyses, the mean In abundance in H5 chondrites is ~1.0 ng/g, which is comparable to In abundances in H4 and H6 chondrites. The overall In abundance trend is broadly similar to that of Cd and Ag in H-chondrites (H3 > H4 – H6). The mean In abundances decrease from L3 to L6 chondrites, but the In abundances in L3 chondrites do not show a trend with petrologic type. There are correlations between abundances of In and the noble gases Ar and Xe (Keays et al., 1971) and of In and carbon (Tandon and Wasson, 1968) in L-chondrites. In contrast to the H- and L-chondrites, the In abundances in LL-chondrites do not vary systematically with petrologic type because the LL4 chondrites are out of sequence (LL3 > LL5 > LL6 > LL4). However, In abundances in LL3 chondrites are significantly higher than those in LL4 – LL6 chondrites.

In meteorites, thallium is dispersed primarily in troilite and also substitutes for K in feldspars (Mason, 1971). Thallium abundances (like those of In, Cd, and Ag) are significantly larger in H3 than in H4 – H6 chondrites. The median Tl abundances for H4 – H6 chondrites are essentially the same, but the mean Tl abundances increase with increasing petrologic type (H4 < H5 < H6). The same behavior was observed for Tl by Wolf and Lipschutz (1998), although they did not consider type 3 chondrites. Within the H3 subtype, Tl abundances show a similar trend to In and decrease from H3.4 to H3.9. Given the similarity to In and Cd, we believe that the median Tl abundances, rather than the means, are more representative of actual trends within the H4 – H6 chondrites. In L-chondrites, Tl shows broadly similar behavior to Cd. This is particularly true if we neglect one high analysis for the Rupota L4-6 chondrite (119 ng/g). Doing so lowers the mean Tl abundance in L4 chondrites to ~6 ng/g, which is only slightly larger than the In abundance in L3 chondrites. Thallium abundances then vary in the sequence L3 ~ L4 > L5 ~ L6, as seen with Cd, although the distinction is not quite as sharp. The LL3 chondrites also have significantly higher Tl concentrations than higher petrologic types. The median Tl abundances increase from LL6 to LL5 to LL4 chondrites, but the mean abundances, which are significantly different for LL5 and LL6 types, do not show the same trend. However, this behavior is again broadly similar to that of Cd in LL-chondrites. The correlation between In, Tl, Cd, and Ag indicates that their chemistry within the meteorites should be similar.

*3.1.5. Group IVA: C, Ge, Sn, Pb.* Schaefer and Fegley (2007a) discussed Si, and we consider C, Ge, Sn, Pb below. Carbon abundances are shown in Fig. 5. As can be seen, the abundance of carbon in H3 chondrites is significantly higher than in types H4-H6, which have relatively constant abundances. For L-chondrites, the abundance of carbon in L3s and L4s is about the same, and then there is a slight decrease in abundance for the L5 and L6 chondrites. There is limited data on carbon abundances in LL-chondrites, but it appears that carbon is significantly more abundant in LL3 chondrites than in LL4 – LL6 chondrites, which have similar abundances.

The abundance of Ge in H-, L-, and LL- chondrites with petrologic type is also shown in Fig. 5. Germanium is found almost exclusively within the metal phase of ordinary chondrites; however, there is more Ge in silicates in type 3 ordinary chondrites than in the higher petrologic types (Chou et al., 1973). Germanium abundances in equilibrated H-chondrites increase from H6 to H4, but Ge abundances are lowest in the H3 chondrites, for which there are only two analyses (both of H/L3 chondrites). In L-chondrites, Ge abundances increase from L3 to L5; however, Ge abundances in L6 chondrites are essentially the same as in L3 chondrites. There are no Ge analyses for LL4 or LL6 chondrites so it is unclear if the Ge content varies with petrologic type in LL chondrites.

Tin is concentrated primarily in the metallic phases in chondritic meteorites. Tin analyses are subject to analytical problems (Mason, 1971). We have data only for H3, H5, L3, L6, and LL3 chondrites, which are not enough to determine trends with petrologic type. Lead is a dispersed (highly variable) element found primarily in troilite. One study of Pb distribution in the Khohar L3 chondrite found a slight enrichment within kamacite grains, but the Pb was also fairly well-distributed throughout the matrix and sulfide phases (Woolum and Burnett, 1981). There are only a few Pb elemental analyses for ordinary chondrites in the literature. Additionally, the Pb in chondrites is subject to significant contamination from terrestrial sources, so the indigenous abundance of Pb is difficult to determine (Unruh, 1982). We only have data for H4 – H6, L5, L6, and LL6 chondrites and cannot determine if trends exist with petrologic type for the L- and LL-chondrites. The mean Pb abundances in H-chondrites increase from H4 – H6 but the median Pb abundances are essentially constant. We have no data for Pb in H3 chondrites.

*3.1.6. Group VA: As, Sb, Bi.* Abundances of As, Sb, and Bi in H-, L-, and LL-chondrites are shown in Fig. 6. Arsenic is primarily siderophile in meteorites, but there is also substantial As in troilite (Rambaldi et al., 1978). The range of As abundances in H3 chondrites spans the entire range of all petrologic types for H-chondrites. The mean and median As abundances increase from H3 – H5, however the As abundance in H6 chondrites is lower than that of H5 chondrites. A very similar trend is observed in the L-chondrites, although the overall abundance of As is lower in the L-chondrites than in H-chondrites because the L-chondrites contain less Fe metal than H-chondrites. The As abundance in LL-chondrites is fairly constant (LL3 – LL6), although its mean and median abundances in LL5 chondrites are lower than that for the other petrologic types.

Antimony is siderophile and chalcophile and so is expected to concentrate within the metallic and sulfide phases of meteorites (Mason, 1971). Antimony abundances for H- and L-chondrites are fairly constant across petrologic type. The very large range for Sb in H4 chondrites is due to one low value for Cañellas (1.5 ng/g). If we discard this analysis, the range is comparable to that for Sb in H5 and H6 chondrites, and the mean and median abundances increase to values roughly equal to that of H3 chondrites. The large range in L6 chondrites is due to one large analysis for Tourinnes-la-Grosse (840 ng/g). If we reject this analysis, the mean Sb abundance becomes essentially the same as that in L5 chondrites. The Sb abundance in LL-chondrites scatters with LL4 > LL3 > LL6 ~ LL5.

Bismuth is primarily chalcophile and may reside in troilite (Mason, 1971). However, one study of the Khohar L3 chondrite found that Bi was highly localized in Ni-poor kamacite (Woolum and Burnett, 1981). Bismuth abundances show similar trends to In, Tl, and, to a lesser extent, Cd and Ag. The Bi abundance of unequilibrated H3 chondrites decreases from H3.4 to H3.9. The Bi abundance in H3 chondrites is much higher than that of H4 – H6 chondrites, which have essentially constant Bi contents. In L-chondrites, the Bi abundance decreases with increasing petrologic type from L3 to L6. The large range for L6 chondrites is due to a single large value for Milena (516 ng/g); if we remove this analysis, the mean Bi abundance of L6 chondrites drops to ~5 ng/g, in much better agreement with the median abundance. The LL3 chondrites have higher Bi abundances than LL4 – LL6 chondrites, which show increasing Bi abundances from LL4 to LL6.

*3.1.7. Group VIA: Se, Te.* The abundances of Se and Te in H-, L-, and LL-chondrites are shown in Figure 7. Most of the Se in ordinary chondrites is in troilite (Curtis and Schmitt, 1979). Most of the sulfur in ordinary chondrites is in troilite, so the two elements should correlate with one another. We combined the mean Se abundances in Tables 1-3 and the mean S abundances for H-, L-, LL-chondrites from Table 1 of Schaefer and Fegley (2007a) to calculate the average S/Se ratios in ordinary chondrites. Our results give S/Se ratios of 2371, 2375, and 2433 in H-, L-, and LL-chondrites, respectively, and an overall mean of 2393 ± 37 ($1\sigma$). This shows that the Se/S ratios of ordinary chondrites are identical within analytical uncertainties, as was also observed by Dreibus et al. (1995). The Se abundance in H- and L-chondrites is essentially constant across petrologic type. The Se abundances in L3 and L4 chondrites are nearly identical, but the Se abundances in L5 and L6 chondrites are slightly higher. In contrast, the Se abundances in LL3 and LL4 chondrites are larger than those in LL5 and LL6 chondrites. This is very similar to the pattern of Sb abundance in LL-chondrites.

Tellurium is chalcophile and siderophile so Te should be present in sulfide and metal. A comparison of S/Te ratios in H- (55,862), L- (58,112), and LL- (47,472) chondrites shows greater variability than the S/Se ratios and is consistent with some Te being in the metal phase. The median Te abundances in Tables 1-3 were used in these calculations, but a similar pattern emerges using the mean Te values. Mason (1971) reports that Te is very susceptible to leaching, indicating that at least some of it is located in water-soluble phases, perhaps CaS, $FeCl_2$, and MgS. Tellurium abundances show similar trends to Se for H- and LL-chondrites. For H3 chondrites, one very large Te abundance for Sharps H3.4 (3200 ng/g) raises the mean drastically. If we reject this analysis, the mean and median Te abundances are essentially identical for H3 – H6 chondrites. Tellurium abundances in LL-chondrites match the pattern shown by Se, with LL3 and LL4 chondrites containing more Te than LL5 and LL6 chondrites. In contrast to Se, the L3 chondrites have higher Te abundances than L4 – L6 chondrites. If we reject two high values for Barwell L5 (1150 ng/g) and Innisfree L5 (1534 ng/g), the mean and median Te abundances for L4 – L6 chondrites are constant.

*3.2 Correlations of Trace Element Patterns*

*3.2.1 H-chondrites* We found four different abundance trends in H-chondrites. (1) Gold, Rb, Zn, Ga, Te, Sb, Cu, and Se show essentially constant abundances with petrologic type. This agrees with the work of Wolf and Lipschutz (1998), who observed no trends for the elements Rb, Se, Te, or Zn with petrologic type. (2) Bismuth, In, and Tl display similar patterns, with abundances decreasing with increasing petrologic type from 3.4 to 3.9, as shown in Fig. 8(a), and then remaining relatively flat across types 4, 5, and 6. Fig. 8(b) shows the abundance of carbon for H3 and H/L3 chondrites. The carbon content decreases slightly from H3.4 to H3.8, but seems to increase from H/L3.6 to H/L3.9 (although a trend is hard to establish within the H/L group since only two type 3 subtypes are represented). Similarly, Cd and Ag are significantly more abundant in H3 chondrites than in the higher petrologic types. There are only three analyses for Cd in H3 chondrites, which do indicate a decreasing trend with increasing petrologic type, but it is difficult to accept a trend based on only three points. Wolf and Lipschutz (1998) did not observe any patterns for Ag, Cd, Bi, Tl, or In because their data set did not include any H3 chondrites. As we have noted, the abundances for these elements in H4 – H6 chondrites are relatively constant, which agrees with the analysis of Wolf and Lipschutz (1998). (3) Cesium and Ge have similar patterns, with abundances decreasing from H4 – H6 chondrites. Our trend for Cs matches that observed by Wolf and Lipschutz (1998) in H4 – H6 chondrites. This trend does not continue into H3 chondrites, but this may be due to the use of the H/L chondrites Tieschitz and Bremervörde in our compilation. However, the only Ge analyses in unequilibrated ordinary chondrites are for H/L3 chondrites. (4) Arsenic has a pattern distinct from all other trace elements. The As abundance increases from H3 to H5 chondrites and then decreases in H6 chondrites. This pattern is remarkably similar to that for As in L-chondrites.

*3.2.2 L-chondrites* We found five distinct abundance trends in L-chondrites. (1) Gold, Rb, Zn, Ga, Te, Sb, Cu, and Se are essentially constant across petrologic type. We also observed this pattern in H-chondrites. (2) Bismuth, In and Cs have similar patterns, with abundances decreasing with increasing petrologic type from L3 to L6 chondrites. There are no sub-type trends in L3 chondrites as observed for H3 chondrites. (3) Thallium, Cd and C have similar abundances in L3 and L4 chondrites, which are larger than their similar abundances in L5 and L6 chondrites. (4) Silver has a distinct pattern with L3 chondrites containing significantly more Ag than L4 – L6 chondrites, which have lower, relatively constant abundances. This is similar to the pattern for Ag in H- and LL-chondrites. (5) Both As and Ge have similar patterns, with abundances increasing from L3 to L5 chondrites, and then decreasing in L6 chondrites. Arsenic shows a nearly identical pattern in H-chondrites.

*3.2.3 LL-chondrites* We also found five types of patterns in LL-chondrites. (1) Copper, Au, Zn, and possibly As, have essentially constant abundances with petrologic type. (2) Tl, Cd, C, and tentatively In, are more abundant in LL3 chondrites than in LL4 – LL6 chondrites. (3) Cs and Ag show similar patterns, with abundances decreasing from type 3 to type 5, and then increasing in type 6 ($3 > 4 > 5 < 6$). (4) Rb, Ga, and Bi have similar patterns, with abundances increasing from type 4 to type 6, and similar abundances in types 3 and 6 ($4 < 5 < 6 \lesssim 3$). This is strikingly different than the abundance patterns for these elements in H- and L-chondrites. (5) Sb, Se, and Te have similar patterns, with abundances in LL3 and LL4 chondrites being larger than in LL5 and LL6 chondrites ($(3 < 4) > (5 < 6)$). This is broadly similar, but not identical, to the Tl and Cd patterns in L-chondrites.

**4. Methods**

We used chemical equilibrium calculations to model volatile element chemistry during metamorphism of ordinary chondritic material. We did chemical equilibrium calculations using a Gibbs energy minimization code of the type described by Van Zeggern and Storey (1970). Our calculations consider major rock-forming elements (Al, Ca, Fe, K, Mg, Na, O, Si, Ti), minor elements (Co, Cr, Mn, Ni, P, S), trace elements (Cu, Zn, Ga, Ge, As, Se, Rb, Ag, Cd, In, Sn, Sb, Te, Cs, Au, Tl, Pb, Bi), and volatiles (C, Cl, F, H, N, Br, I) in chondrites. We used the IVTANTHERMO database (Belov et al., 1999), to which we added a number of minerals found in chondrites. Approximately 1700 compounds of 40 elements were included in the calculations. Our nominal model uses ideal solid solutions of olivine $(Mg,Fe,Zn,Mn)_2SiO_4$, pyroxene $(Mg,Ca,Fe,Zn,Mn)SiO_3$, feldspar $(Na,K,Ca)_2(Al,Si)_3SiO_8$, and metal (Fe, Ni, Co). For the majority of the trace elements (except Zn), solid phases were considered to be pure compounds. In reality, most trace elements in meteorites typically do not form discrete minerals, but dissolve in more abundant host phases or are dispersed among several phases. However, as discussed earlier, the host phases of several trace elements are unknown. In addition, the thermodynamic properties of trace element solid solutions in the host minerals are generally unknown. Solid solution formation typically raises the temperatures at which 50% of an element is in the gas and 50% in the solid, so our calculated stability temperatures are lower limits in several cases. In other words, elements that dissolve in host phases may be more refractory and less volatile than we calculate. However, none of the elements will be less refractory (i.e., more volatile) than we calculate. The reason for this is simple. Our calculations include all gases for which thermodynamic data are available for each trace element. It is unlikely that we have neglected any important gases because in general, thermodynamic data are available for the important compounds of each element. Thus, our results give the maximum possible volatility for the elements included in our calculations.

We used temperature (300 – 1600 K) and pressure (0.1 – 100 bar) ranges appropriate for P – T profiles within possible meteorite parent bodies (i.e., asteroids). We discuss calculation of P - T profiles for the asteroid *6 Hebe*, a prime candidate for the H – chondrite parent body in our earlier paper (Schaefer and Fegley, 2007a). Lodders (2003) describes thermodynamic calculations for gas – solid chemical equilibria.

We used the average (mean) major element compositions of H-, L-, and LL-chondrites in Table 1 of Schaefer and Fegley (2007a) for our nominal models. We used the median trace element abundances given in Tables 1-3 as the nominal abundances in our model. The mean abundances are typically larger than the median abundances. We chose median abundances because the inclusion of one or two anomalously large analyses often significantly raises mean abundances as discussed above. Where means are more than 50% larger than the medians (bold-faced elements in Tables 1-3), the effect of variable abundance on the chemistry of that element was evaluated separately.

**5. Results**

Table 4 summarizes our results for H-chondritic material. It lists temperatures for the initial formation of solid phases for each of the trace elements considered in our calculations at a total pressure of 1 bar. Two elements (Cu and Zn) are completely condensed at temperatures higher than the highest temperature considered in our calculations. The other elements are completely in the gas phase at higher temperatures. For example, Au is in the gas, primarily as AuS, at temperatures above 1220 K where gold metal, the first Au-bearing solid forms. Likewise, Tl remains in the gas, as TlCl, TlBr, and TlI down to 485 K where solid thallium iodide TlI forms. Table 4 also lists the major gases for each element and compares the calculated solid phase with the residence site(s) of each element in ordinary chondrites.

Figures 9-14 show the pressure dependence of the solid phase formation temperatures. The temperature and pressure profile for the nominal H-chondrite parent body *6 Hebe* (Schaefer and Fegley, 2007a) is shown for comparison. The curves in Figures 9-14 are analogous to condensation curves in the solar nebula. However, the curves in Figures 9-14 are for a system with the average composition of H-chondritic material, instead of for solar composition material. We discuss our results below according to periodic table groups.

*5.1 Formation Temperatures of Solid Phases of the Trace Elements*

*5.1.1 Group IA: Rb, Cs* Formation temperatures for solid phases of Rb and Cs are shown in Fig. 9. As discussed earlier, Rb and Cs are found in water-soluble phases and feldspars in ordinary chondrites (Gast, 1960, 1962; Smales et al., 1964; Curtis and Schmitt, 1979; Wieler et al., 2000). Thermodynamic data for Cs and Rb feldspars are unavailable so we could not include them in our calculations. However, a number of inorganic compounds of Rb and Cs are in the calculations (borates, carbonates, halides, hydrides, hydroxides, nitrates, nitrites, oxides, sulfides, sulfates). We found that Rb and Cs condensed as halide salts. In H-chondrites these are RbBr (s,l), RbCl (s,l), and CsI (s). At a pressure of 1 bar, the formation temperatures for these phases are 860 K, 940 K, and 760 K, respectively. As shown in Figure 9, the formation curves are linear with respect to log pressure and the curves for Rb-salts are more pressure dependent than that for CsI (s). In L-chondrites, RbI (s) also forms at 780 K at one bar pressure. Cesium iodide (s) and CsBr (s) form in LL-chondritic material, with nearly identical curves to that for CsI (s) in H-chondritic material.

The mean abundances of Cs in L- and LL-chondrites deviate significantly from that of the median abundances (6.1 and 1.9 times larger, respectively). When the abundances of Cs in L- and LL-chondritic material are raised to their mean values, the formation temperatures of CsI (s) increase by 39 K and 22 K, respectively. Additionally,

CsBr (s), which is not stable in our nominal L-chondritic material, forms at ~ 730 K, but the formation temperature of CsBr (s) in LL-chondritic material does not change.

As mentioned above we do not have thermodynamic data for Rb and Cs feldspars. However, the observations of Rb in sylvite in the Monahans (1998) H5 chondrite suggest that some Rb may partition into feldspar, possibly via exchange reactions such as

$$RbCl \text{ (in sylvite)} + KAlSi_3O_8 \text{ (in feldspar)} = KCl \text{ (in sylvite)} + RbAlSi_3O_8 \text{ (in feldspar)} \quad (1)$$

Similar reactions can be envisioned for Cs, but the work of Curtis and Schmitt (1979) shows that Cs can be dispersed among several phases in ordinary chondrites.

*5.1.2 Group IB: Cu, Ag, Au* Formation temperatures for solid phases of Cu, Ag, and Au are shown in Fig. 10. In average H-chondritic material, copper forms as chalcocite $Cu_2S$ (s) at temperatures greater than the highest temperatures used in our calculations at all pressures (1600 K). Chalcocite $Cu_2S$ (s) converts into copper metal at 895 K, independent of pressure, via the net thermochemical reaction

$$Cu_2S \text{ (chalcocite)} + H_2 = 2Cu \text{ (metal)} + H_2S \quad (2)$$

In L- and LL-chondritic material, this conversion takes place at temperatures 24 K and 184 K lower, respectively, than in H-chondritic material. Silver and gold condense as metals, and their formation temperatures show a linear dependence with log pressure. In H-chondritic material, Ag metal forms at 1175 K, and Au metal forms at 1220 K (both at 1 bar total pressure). Formation temperatures of Ag (s,l) and Au (s,l) are similar in L- and LL-chondritic material. However, in L-chondritic material the formation curves for Au (s,l) and Ag (s,l) cross at high pressures, so that at the highest pressures, Ag is stable at higher temperatures than Au. The formation temperature of Au (s,l) in LL-chondritic material is nearly identical to that of Ag (s,l).

In average H-chondritic material, the mean Ag abundance (78.2 ng/g) is more than 2 times larger than the median value (36.4 ng/g). In L-chondritic material, the mean Ag abundance (126.61 ng/g) is 1.9 times larger than the median value (65.7 ng/g). The discrepancy is not as large for LL-chondritic material (a factor of 1.2). When the Ag abundances are increased to their median abundances, the formation temperatures of Ag in H- and L-chondritic materials increase by 33 K and 30 K, respectively, at 1 bar pressure.

*5.1.3 Group IIB: Zn, Cd* Figure 11 shows the pressure dependent formation temperatures for solid phases of Zn and Cd. Zinc chemistry is fairly complicated and is significantly different for calculations using pure Zn-bearing phases versus solid solutions. If we consider only pure Zn-bearing phases, at high pressures (> 100 bars), ZnS (wurtzite) is the first condensate, which is converted at lower temperatures to ZnS (sphalerite). At still lower temperatures, ZnSe (stilleite) and then ZnTe also condense, but they do not completely consume ZnS (sphalerite) because Se and Te are less abundant than zinc.

When zinc is allowed to form silicate solid solutions, zinc dissolves in forsterite as $Zn_2SiO_4$ (willemite) and in enstatite as $ZnSiO_3$ at temperatures greater than our peak calculation temperature. Stilleite ZnSe forms at 830 K at a pressure of 1 bar, and its formation temperature varies linearly with log pressure. Zinc chromite $ZnCr_2O_4$ forms at a temperature of ~650 K, independent of pressure. Zinc sulfide and ZnTe do not form when Zn-bearing solid solutions are considered. Formation temperatures for Zn-bearing phases in H-, L-, and LL- chondritic materials are essentially identical.

Zinc chemistry affects cadmium chemistry through Se and Te. When only pure Zn-bearing phases are allowed, CdTe (s) forms at ~760 K (1 bar pressure), and its formation temperature varies linearly with log pressure. When Zn-silicate solid solutions form, the initial Cd condensate is CdTe (s) at pressures less than ~1 bar and CdSe (s) at higher pressures. At all pressures, CdSe converts into CdTe (s) at ~730 K and forms from it again at ~ 470 K, independent of pressure, via a net reaction such as

$$CdSe\ (s) + H_2Te\ (g) = CdTe\ (s) + H_2Se\ (g) \qquad (3)$$

Formation temperatures of Cd-bearing solids in H-, L-, and LL-chondritic materials are essentially identical.

In average H-chondritic material, the mean Cd abundance (37.1 ng/g) is more than 8 times larger than the median value (4.59 ng/g). In L-chondritic material, the mean Cd abundance (61.4 ng/g) is 2.4 times larger than the median value (25.95 ng/g). In LL-chondritic material, the mean Cd abundance (76.18 ng/g) is ~4.5 times larger than the median value (17 ng/g). When the Cd abundances are increased to their median abundances, the initial formation temperatures of CdSe in H-, L-, and LL-chondritic materials increase by 66 K, 28 K, and 41 K, respectively, at 1 bar pressure. The formation temperature of CdTe does not change when the abundance of Cd is raised because it forms from pre-existing CdSe and its formation temperature depends only on the $Se_2/Te_2$ partial pressure ratio and the equilibrium constant for the reaction above.

*5.1.4 Group IIIA: Ga, In, Tl* Formation temperatures for solid phases of Ga, In, and Tl are shown in Fig. 12. Gallium condenses into $Ga_2O_3$ (s) at high temperatures. The one bar formation temperature in H-chondritic material is ~1140 K. Formation temperatures for $Ga_2O_3$ in L- and LL-chondritic material are essentially identical to that in H-chondritic material. Indium and thallium are much more volatile than Ga and condense at lower temperatures as InS (s) and TlI (s), respectively. Their 1 bar formation temperatures in H-chondritic material are 503 and 484 K, respectively. In H-chondritic material, the formation temperature of TlI (s) is slightly more pressure dependent than that of InS, so at high pressures (>150 bars), TlI forms at higher temperatures. Formation curves for InS in L- and LL-chondritic material are slightly higher than that in H-chondritic material while the TlI formation curves are essentially identical. For L- and LL-chondritic material, the formation temperature of InS is always higher than TlI.

Mean and median abundances of In and Tl are significantly different in all three types of ordinary chondrites. For In, the mean abundances are 7.2, 6.9, and 2.8 times larger than the median abundances in H-, L-, and LL-chondritic material, respectively. For Tl, the mean abundances are 9.3, 2.9, and ~2 times larger than the median abundances in H-, L-, and LL-chondritic material, respectively. When the In abundance is increased to its mean value in H-, L-, and LL-chondritic material, the formation temperature of InS increases by 36 K, 31 K, and 17 K, respectively. When the Tl abundance is increased to its mean value in H-, L-, and LL-chondritic material, the formation temperature of TlI increases by 32 K, ~16 K, and 33 K, respectively.

*5.1.5 Group IVA: Ge, Sn, Pb* Formation temperatures for solid phases of Ge, Sn, and Pb are shown in Fig. 13. Germanium and tin condense into metal initially and have nearly identical slopes as a function of log pressure, with Sn slightly lower than Ge. In H-chondritic material at a pressure of 1 bar, Ge (s) forms at 850 K and Sn (s) forms at 820 K. At lower temperatures, Sn (s) converts into herzenbergite SnS at a temperature of ~600 K, independent of pressure, via the net thermochemical reaction

$$\text{Sn (metal)} + H_2S \text{ (g)} = \text{SnS (herzenbergite)} + H_2 \text{ (g)} \quad (4)$$

At still lower temperatures Ge (s) converts into $GeO_2$ (s) at a temperature of 330 K, independent of pressure, via the net thermochemical reaction

$$\text{Ge (metal)} + 2\ H_2O \text{ (g)} = GeO_2 \text{ (s)} + 2\ H_2 \text{ (g)} \quad (5)$$

Lead condenses entirely as PbSe (s). The formation temperature of PbSe is similar to that of Ge and Sn at low pressures, but is less pressure dependent than the formation curves for these two metals. At a pressure of 1 bar, the formation temperature of PbSe (s) is ~860 K. Formation of PbSe in L- and LL- chondritic material follows similar formation curves, but temperatures are shifted slightly downward by ~5-10 K with shallower slopes. The stability regions of Sn (l) and Ge (s) are smaller in L- and LL-chondritic materials than in H-chondritic material, i.e., they form at lower temperatures. At one bar pressure the differences are ~5 K (L) and ~40 K (LL). Metallic Sn and Ge convert into SnS (s) and $GeO_2$ (s) at higher temperatures than in H-chondritic material. At one bar the temperature differences for SnS are ~10 K (L), ~100 K (LL); for $GeO_2$ the differences are ~18 K (L) and ~ 43K (LL).

*5.1.6 Group VA: As, Sb, Bi* Formation temperatures for solid phases of As, Sb, and Bi are shown in Fig. 14. In average H-chondritic material, we found that arsenic condenses as the metal phase at ~595 K at a pressure of 1 bar. Antimony condenses as a metal at ~750 K. If no zinc solid solutions form, Sb (s) converts into $Sb_2Te_3$ (s) at ~490 K. The appearance of $Sb_2Te_3$ (s) is essentially independent of pressure due to the net reaction

$$2\ \text{Sb (s)} + 3\ H_2Te \text{ (g)} = Sb_2Te_3 \text{ (s)} + 3\ H_2 \text{ (g)} \quad (6)$$

Bismuth condenses as the metallic element, which occurs at ~735 K at 1 bar pressure. If zinc solid solutions form, bismuth metal converts into $Bi_2Te_3$ at lower temperatures independent of pressure (~670 K) due to the net reaction

$$2\ Bi\ (s) + 3\ H_2Te\ (g) = Bi_2Te_3\ (s) + 3\ H_2\ (g) \quad (7)$$

The formation curves for As, Sb, and Bi in L- and LL-chondritic material are nearly identical to their formation curves in H-chondritic material.

The mean abundances of Bi in H-, L-, and LL-chondritic material are significantly larger than the median values, by 3.5, 4.7, and 1.5 times, respectively. Raising the abundance of Bi to the mean values results in formation temperatures of Bi metal that are 28 K, 36 K, and 5 K higher, respectively.

*5.1.7 Group VIA: Se, Te* Results for sulfur were discussed in Schaefer and Fegley (2007a). Formation curves for Se and Te compounds are shown in Fig. 11. In average H-chondritic material, we found that Se initially condenses as ZnSe at a temperature of ~825 K at 1 bar pressure. At pressures higher than ~10 bars, the initial Se solid phase is CdSe. The CdSe formation curve has a linear dependence on log pressure, but CdSe does not form at pressures below ~1 bar. At all higher pressures, CdSe converts into CdTe at a temperature of ~730 K, independent of pressure. CdTe is then reconverted into CdSe at ~470 K, independent of pressure. Iron selenide forms at ~370 K via the net reaction

$$ZnSe\ (s) + FeCr_2O_4\ (chromite) = FeSe\ (s) + ZnCr_2O_4\ (s) \quad (8)$$

The formation of iron selenide destroys ZnSe. The initial Te phase is $FeTe_{0.9}$ (s) at all pressures. Its stability temperature at 1 bar is ~780 K. CdTe forms at lower temperatures. At pressures greater than ~1 bar, the formation temperature of CdTe is ~730 K, independent of pressure. At lower pressures, the formation temperature of CdTe is pressure dependent, with a similar slope to that of $FeTe_{0.9}$ (s).

## 6. Discussion

### *6.1 Volatility sequences*

Table 5 lists the volatility sequences for the trace elements in order of increasing volatility for H-, L-, and LL-chondritic material. The volatility sequences are derived from our results described above. Increasing volatility corresponds to decreasing formation temperatures for solid phases containing an element. Copper and Zn, which are already condensed above 1600 K (see Table 4), are the least volatile (or most refractory) elements in H-chondritic material while Tl, which condenses into solid Tl iodide at 485 K, is the most volatile (or least refractory) element. To our knowledge, there are no other volatility sequences available for ordinary chondritic material. Fegley (1990) modeled volatile trace element transport during ordinary chondrite metamorphism, but only presented results for lead. Previous papers about the volatility of trace elements in ordinary chondritic material relied on the volatilities of these elements in the solar nebula (e.g. Larimer, 1967; Larimer and Anders, 1967). The solar nebula volatility (condensation) sequence calculated by Lodders (2003) is also listed in Table 5 for comparison. Lipschutz and colleagues measured the mobility of trace elements in heated chondritic material, and we discuss their work later.

As can be seen from Table 5, our calculated volatility sequences for trace elements in H-, L-, and LL-chondritic material are very similar to one another with a few minor variations. Cadmium is more volatile in H-chondritic material than in L- or LL-chondritic material, while Cs is more volatile in L-chondritic material. However, the volatility sequences of the ordinary chondritic material are extremely different than that of the solar nebula. In particular, the volatilities of As, Sb, and, to a lesser degree, Bi in ordinary chondritic material are much higher than in the solar nebula. We also find that the solar nebula volatilities of Zn, Cd, and Se are significantly higher than for the ordinary chondrites. Differences in volatility between the solar nebula and ordinary chondrite parent bodies are, however, not too surprising, considering the different environments and compositions involved.

Solar nebular volatilities are calculated assuming total pressures of $10^{-3}$ to $10^{-4}$ bars, whereas much higher lithostatic pressures (1-1,000 bars) are found within ordinary chondrite parent bodies. The solar nebula composition is also significantly more reducing, i.e., has a lower oxygen fugacity, at a given temperature than ordinary chondritic material. The ordinary chondritic compositions are much more oxidizing, which alters the major gases and stable solid phases of the trace elements, and therefore the formation temperatures for the solid phases. As stated by Keays et al. (1971): "It is virtually certain that the condensation curves of [trace elements in the solar nebula] cannot be applied to metamorphic fractionations."

Table 5 also lists a volatile mobility sequence for the unequilibrated ordinary chondrite Tieschitz (H/L3.6) as measured by Ikramuddin et al. (1977). Ikramuddin et al. (1977) heated samples of Tieschitz in a stepped fashion in an atmosphere of $10^{-5}$ atm $H_2$ for temperatures of 400˚ – 1000˚C and measured the amounts of volatile elements retained after heating. Nearly all Cs, Bi, Tl, Ag, In, Te, and Zn are lost from the samples at the end of the heating run, whereas most Ga and Se remain. The mobility of thermally labile elements is governed more by kinetic factors such as diffusion and desorption than by thermodynamics (Ikramuddin et al. 1977), making comparisons to our calculated volatility sequences problematic. However, the comparison is an instructive one, showing that elements that we calculate to be stable within an ordinary chondrite parent body, such as Zn, Cs, and Ag, may, in fact, move around during open system metamorphism.

We did calculations for conditions comparable to the experimental conditions of Ikramuddin et al. (1977) ($P \sim 10^{-5}$ bar, T=300-1300 K), in order to determine if the differences between the calculated volatilities and the observed mobility sequence are due to differences in pressure. At the lower pressure conditions, we found that Ag and Zn became significantly more volatile, whereas the volatility of the other elements considered by Ikramuddin et al. (1977) did not change considerably. The initial formation temperature for Ag metal dropped from 1176 K at 1 bar to 840 K at $10^{-5}$ bar, and the 50% formation temperature of Zn (as $ZnSiO_3$ in enstatite solid solution) dropped from ~1500 K at 1 bar to ~850 K at $10^{-5}$ bar.

*6.2 Comparison to Trace Element Patterns and Implications for Parent Body Metamorphism*

Going back to our previous discussion on trace element abundance patterns, we can see that there are significant differences between the H- and L- chondrites on the one hand and the LL-chondrites on the other hand. However, our results indicate that there is no significant difference in chemistry for the three different types, so these differences must be due to a different series of primary and secondary processes operating within the parent body.

From our model, we predict that the elements Rb, Cu, Au, Ag, Zn, and Ga should be unaffected by thermal metamorphism within an H-chondrite parent body similar to *6*

*Hebe* due to their low volatility and high formation temperatures of their solid phases. More volatile elements should be more easily vaporized and transmitted throughout the parent body during thermal metamorphism. These are the elements that should then have highly variable abundances with petrologic type. According to our calculations, these elements include Tl, In, As, Sb, Bi, Cs, Te, and Cd, whereas Sn, Pb, Ge, and Se should be only moderately affected by metamorphism.

If we compare our predictions to the observed abundance patterns in H- and L-chondrites, we can see that, while some agree, there are several striking discrepancies. In particular, Ag, which we calculate to be relatively refractory, is highly enriched in the type 3 chondrites, similar to the behavior of Bi, In, and Tl. This is in much better agreement with the measured mobility sequence of Ikramuddin et al. (1977). If we assume a thermal metamorphism mechanism, Ag can be enriched in the type 3s by mobilizing the Ag from the interior and re-depositing it in the outer regions of the parent body, along with Bi, In, and Tl.

Another disparity between our predicted results and the observed abundance patterns is for As and Sb. Our calculations predict that these elements should behave in a similar fashion to Bi, with As actually being more volatile than Bi. However, whereas Bi is enriched in type 3 chondrites, As and Sb have relatively constant abundances across all petrologic types. Solid solution of As and Sb in iron metal may decrease their volatility, but Bi should also be found in metal solid solutions, so it is difficult to see how this would lead to such a disparity.

It seems that the mobility sequence measured by Ikramuddin et al. (1977) matches the observed abundance patterns in H- and L-chondrites better than the calculated volatilities do. However, the correlation between mobility sequence and abundance pattern is not perfect because Te and Zn, which have relatively constant abundances in the H- and L- chondrites, are listed amongst the highly mobile elements. In fact, Te should be more mobile than Ag, but this is clearly not the case.

LL-chondrite abundance patterns cannot be explained using either element mobility or volatility. There are significantly fewer analyses of trace elements in LL-chondrites than in the other ordinary chondrite classes. It may be that the patterns we see now in the trace elements are simply due to the small sampling size, and that more normal (i.e., similar to the H- and L- chondrites) abundance patterns would emerge with more data. However, it seems at this stage more likely that there were simply different processes operating within the LL-chondrite parent body than in the H- and L- chondrite parent bodies. It is also possible that while the onion-shell model (type 3 through type 6 increase with depth inside the parent body) is likely true for the H- and L- chondrites, the LL-chondrite parent body may be of the rubble-pile type (that is, the parent body was collisionally fragmented during heating and then reassembled over a short time span) (Grimm, 1985). In a rubble-pile parent body, all petrologic types would be jumbled together while still fairly hot, so systematic trends would not be likely for trace element abundances.

**7. Summary**

Abundance trends observed within H- and L- chondrite parent bodies support the theory of trace element mobilization in an onion-shell type parent body. However, element mobilities are significantly different than their predicted volatilities at the possible pressure and temperature conditions within a parent body. Abundance trends within LL-chondrites are significantly different than in the H- and L-chondrites, and do not support systematic variation with petrologic type. Rather, they seem to indicate that the LL-chondrite parent body experienced a much more complex thermal history than the H- and L-chondrites, and may support a rubble-pile model for the LL-chondrite parent body.

## 8. Acknowledgements

We acknowledge support from the NASA Astrobiology and Outer Planets Research Programs and the McDonnell Center for the Space Sciences. We thank J. Friedrich, K. Lodders, and S. Wolf for helpful discussions.

**Table 1.** H-chondrite trace element abundances

| Element | Range | Mean ± 1σ | Median | No. of analyses |
|---|---|---|---|---|
| Cu (μg/g) | 48 - 137 | 98 ± 20 | 97 | 51 |
| Zn (μg/g) | 4.6 – 144 | 48 ± 16 | 45.95 | 150 |
| Ga (μg/g) | 3.6 - 11 | 6.0 ± 1.1 | 5.8 | 132 |
| Ge (μg/g) | 5.04 - 16 | 9.4 ± 3.4 | 9.86 | 21 |
| As (μg/g) | 0.35 – 6.71 | 2.6 ± 1.3 | 2.18 | 64 |
| Se (μg/g) | 4.32 - 20.6 | 8.2 ± 1.4 | 8.18 | 157 |
| Br (μg/g) | 0.017 - 1.30 | 0.5 ± 0.4 | 0.34 | 22 |
| Rb (μg/g) | 0.43 - 4.0 | 2.2 ± 0.8 | 2.18 | 143 |
| **Ag (ng/g)** | 4.69 - 1870 | 78 ± 178 | 36.4 | 129 |
| **Cd (ng/g)** | 0.15 - 1240 | 37 ± 153 | 4.59 | 112 |
| **In (ng/g)** | 0.05 - 103 | 3.6 ± 12 | 0.49 | 116 |
| **Sn (μg/g)** | 0.24 - 1.1 | 0.58 ± 0.46 | 0.39 | 3 |
| Sb (ng/g) | 1.5 - 480 | 93 ± 72 | 71.0 | 133 |
| Te (ng/g) | 17 - 3200 | 376 ± 289 | 348 | 116 |
| I (ng/g) | -- | -- | 60 | 1 |
| Cs (ng/g) | 1.29 - 556 | 70 ± 70. | 61.25 | 136 |
| Au (ng/g) | 116 - 440 | 222 ± 51. | 216 | 111 |
| **Tl (ng/g)** | 0.069 - 220 | 8.4 ± 26. | 0.9 | 107 |
| Pb (μg/g) | 0.031 - 0.66 | 0.16 ± 0.17 | 0.14 | 13 |
| **Bi (ng/g)** | 0.14 - 76.8 | 9.2 ± 18. | 2.61 | 124 |

**Table 2.** L-chondrite trace element abundances

| Element | Range | Mean±1σ | Median | No. of analyses |
|---|---|---|---|---|
| Cu (μg/g) | 64 - 132 | 98 ± 18 | 94.65 | 72 |
| Zn (μg/g) | 17.3 - 291 | 57 ± 29 | 52 | 121 |
| Ga (μg/g) | 2.8 - 9 | 5.4 ± 0.8 | 5.5 | 137 |
| Ge (μg/g) | 8.9 - 13.4 | 11 ± 1 | 10.7 | 19 |
| As (μg/g) | 0.32 - 4.2 | 1.4 ± 0.7 | 1.35 | 44 |
| Se (μg/g) | 3.5 - 13.4 | 9.2 ± 1.7 | 9.24 | 93 |
| Br (μg/g) | 0.04 - 5.57 | 1.1 ± 1.2 | 0.89 | 22 |
| Rb (μg/g) | 0.58 - 7.20 | 2.5 ± 0.8 | 2.32 | 106 |
| **Ag (ng/g)** | 14.4 - 1258 | 127 ± 215 | 65.7 | 81 |
| **Cd (ng/g)** | 0.4 - 875 | 61 ± 122 | 25.95 | 66 |
| **In (ng/g)** | 0.03 - 55 | 3 ± 8 | 0.5 | 92 |
| Sn (μg/g) | 0.28 - 1.16 | 0.65 ± 0.37 | 0.59 | 4 |
| Sb (ng/g) | 27.4 - 840 | 103 ± 107 | 76.25 | 72 |
| Te (ng/g) | 153 - 1534 | 405 ± 188 | 376 | 84 |
| I (ng/g) | -- | -- | 70 | 1 |
| **Cs (ng/g)** | 1.64 - 2270 | 82 ± 257 | 13.4 | 91 |
| Au (ng/g) | 27.9 - 339 | 165 ± 44 | 161 | 118 |
| **Tl (ng/g)** | 0.03 - 119 | 4 ± 14 | 1.45 | 79 |
| **Pb (μg/g)** | 0.022 - 0.56 | 0.16 ± 0.19 | 0.063 | 7 |
| **Bi (ng/g)** | 0.09 - 519 | 17 ± 63 | 3.63 | 77 |

**Table 3.** LL-chondrite trace element abundances

| Element | Range | Mean±1σ | Median | No. of analyses |
|---|---|---|---|---|
| Cu (µg/g) | 60 - 147 | 92 ± 22 | 86 | 32 |
| Zn (µg/g) | 28 - 102 | 55 ± 14 | 55 | 61 |
| Ga (µg/g) | 3.3 - 8 | 5.1 ± 1.0 | 5.1 | 55 |
| Ge (µg/g) | 0.2 - 11.2 | 8.1 ± 4.5 | 9.7 | 5 |
| As (µg/g) | 0.08 - 2 | 1.2 ± 0.4 | 1.3 | 32 |
| Se (µg/g) | 4.1 - 18.4 | 8.8 ± 2.6 | 8.6 | 50 |
| Br (µg/g) | 0.2 - 1.98 | 0.93 ± 0.53 | 0.8 | 17 |
| Rb (µg/g) | 0.1 - 5.5 | 2.1 ± 1.2 | 1.95 | 34 |
| Ag (ng/g) | 12.3 - 356 | 80 ± 63 | 64.7 | 32 |
| **Cd (ng/g)** | 0.62 - 1248 | 76 ± 236 | 17 | 29 |
| **In (ng/g)** | 0.09 - 81 | 11 ± 21 | 4 | 38 |
| Sn (µg/g) | -- | -- | 0.33 | 1 |
| Sb (ng/g) | 34.5 - 120 | 74 ± 24 | 69.5 | 31 |
| **Te (ng/g)** | 96 - 4300 | 819 ± 960 | 451 | 32 |
| I (ng/g) | -- | -- | | 1 |
| **Cs (ng/g)** | 1.43 - 3070 | 219 ± 523 | 118 | 33 |
| Au (ng/g) | 97.5 - 306 | 143 ± 36 | 142 | 49 |
| **Tl (ng/g)** | 0.07 - 114 | 15 ± 27 | 1.735 | 42 |
| **Pb (µg/g)** | 0.058 - 0.35 | 0.16 ± 0.17 | 0.058 | 3 |
| **Bi (ng/g)** | 0.57 - 66 | 18 ± 17 | 11.7 | 46 |

**Table 4.** Formation Temperatures of Trace Elements in an Average H-chondrite[a]

| element | $T_{form}$(K) | Initial Solid Phase | Major Gases | Meteorite residence minerals |
|---|---|---|---|---|
| Cu | >1600 | $Cu_2S$ | Cu | metal, troilite |
| Zn | >1600 | Zn-silicates | Zn | silicates |
| Ga | 1140 | $Ga_2O_3$ | GaCl | silicates and oxides |
| Ge | 850 | Ge | GeS, GeSe | metal |
| As | 595 | As | $As_4$, $As_2$, $AsH_3$ | metal, sulfide |
| Se | 860 | PbSe | GeSe, $H_2Se$ | sulfide (troilite) |
| Rb | 865 | RbCl | RbCl, RbBr | silicates |
| Ag | 1175 | Ag | Ag, AgBr | metal, troilite |
| Cd | 735 | CdTe | Cd | troilite (?) |
| In | 505 | InS | InCl, InBr, InI | non-magnetic fraction |
| Sn | 820 | Sn(l) | SnSe, SnS, SnTe | metal |
| Sb | 750 | Sb | $Sb_4$, $Sb_2$, SbTe | metal, sulfide |
| Te | 780 | $FeTe_{0.9}$ | GeTe, SnTe | sulfide |
| Cs | 760 | CsI | CsCl, CsBr, CsI | silicates |
| Au | 1220 | Au | AuS | metal |
| Tl | 485 | TlI | TlCl, TlBr, TlI | troilite |
| Pb | 860 | PbSe | PbTe, PbSe | troilite |
| Bi | 725 | Bi | Bi, BiTe | troilite |

[a]at 1 bar total pressure

**Table 5.** Volatility Sequences in Ordinary Chondritic Material

| | Increasing volatility→ |
|---|---|
| H | Cu, Zn, Au, Ag, Ga, Rb, Pb, Se, Ge, Sn, Te, Cs, Sb, *Cd*, Bi, As, In, Tl |
| L | Cu, Zn, Au, Ag, Ga, Rb, Ge, Pb, Se, Sn, Cd, Te, Sb, Bi, *Cs*, As, In, Tl |
| LL | Cu, Zn, Au, Ag, Ga, Rb, Pb, Se, Ge, Cd, Sn, Cs, Te, Sb, Bi, As, In, Tl |
| Solar[a] | As, Au, Cu, Ag, Sb, Ga, Ge, Rb, Cs, Bi, Pb, Zn, Te, Sn, Se, Cd, In, Tl |
| Mobility[b] | Ga, Se, Cs, Zn, Ag, In, Te, Tl, Bi |

[a]Lodders (2003) [b]Tieschitz (H/L3.6) – Ikramuddin et al. (1977).

**Figure Captions**

Figure 1. Abundances of trace elements in Group IA: (a) Rb, (b) Cs in H-, L-, and LL-chondrite falls. Solid circles and squares represent means and medians, respectively. The bars represent abundance ranges for petrographic types 3, 4, 5, and 6 from left to right.

Figure 2. Abundances of trace elements in Group IB: (a) Cu, (b) Ag, (c) Au in H-, L-, and LL-chondrite falls. Points and squares represent means and medians, respectively. The bars represent petrographic types 3, 4, 5, and 6 from left to right.

Figure 3. Abundances of trace elements in Group IIB: (a) Zn, (b) Cd in H-, L-, and LL-chondrite falls. Points and squares represent means and medians, respectively. The bars represent petrographic types 3, 4, 5, and 6 from left to right.

Figure 4. Abundances of trace elements in Group IIIA: (a) Ga, (b) In, (c) Tl in H-, L-, and LL-chondrite falls. Points and squares represent means and medians, respectively. The bars represent petrographic types 3, 4, 5, and 6 from left to right.

Figure 5. Abundance of trace elements in Group IVA: (a) C, (b) Ge in H-, L-, and LL-chondrite falls. (Sn and Pb are not shown due to low numbers of analyses). There are no analyses of Ge in LL4 or LL6 falls. Points and squares represent means and medians, respectively. The bars represent petrographic types 3, 4, 5, and 6 from left to right.

Figure 6. Abundances of trace elements in Group VA: (a) As, (b) Sb, (c) Bi in H-, L-, and LL-chondrite falls. Points and squares represent means and medians, respectively. The bars represent petrographic types 3, 4, 5, and 6 from left to right.

Figure 7. Abundances of trace elements in Group VIA: (a) Se, (b) Te in H-, L-, and LL-chondrite falls. Points and squares represent means and medians, respectively. The bars represent petrographic types 3, 4, 5, and 6 from left to right.

Figure 8. Abundances of (a) In, Tl, and Bi, and (b) C as a function of petrographic type in H3 and H/L3 chondrites. Solid dots are H3 chondrites, hollow dots are H/L3 chondrites.

Figure 9. Stability temperatures of solid phases for Group IA elements (Rb, Cs) in H-chondritic material. The dotted line shows the temperature-pressure profile of the nominal H-chondrite parent body *6 Hebe* (Schaefer and Fegley, 2007a).

Figure 10. Stability temperatures of solid phases for Group IB (Cu, Ag, Au) elements in H-chondritic material. The dotted line shows the temperature-pressure profile of the nominal H-chondrite parent body *6 Hebe* (Schaefer and Fegley, 2007a).

Figure 11. Stability temperatures of solid phases for Group IIB (Zn, Cd) and Group VIA (Se, Te) elements in H-chondritic material. Zinc is also present in enstatite and olivine solid solutions. The stability temperatures of these phases are higher than our peak temperature. The dotted line shows the temperature-pressure profile of the nominal H-chondrite parent body *6 Hebe* (Schaefer and Fegley, 2007a).

Figure 12. Stability temperatures of solid phases for Group IIIA (Ga, In, Tl) elements in H-chondritic material. The dotted line shows the temperature-pressure profile of the nominal H-chondrite parent body *6 Hebe* (Schaefer and Fegley, 2007a).

Figure 13. Stability temperatures of solid phases for Group IVA (Ge, Sn, Pb) elements in H-chondritic material. The dotted line shows the temperature-pressure profile of the nominal H-chondrite parent body *6 Hebe* (Schaefer and Fegley, 2007a).

Figure 14. Stability temperature of solid phases for Group VA (As, Sb, Bi) elements in H-chondritic material. The dotted line shows the temperature-pressure profile of the nominal H-chondrite parent body *6 Hebe* (Schaefer and Fegley, 2007a).

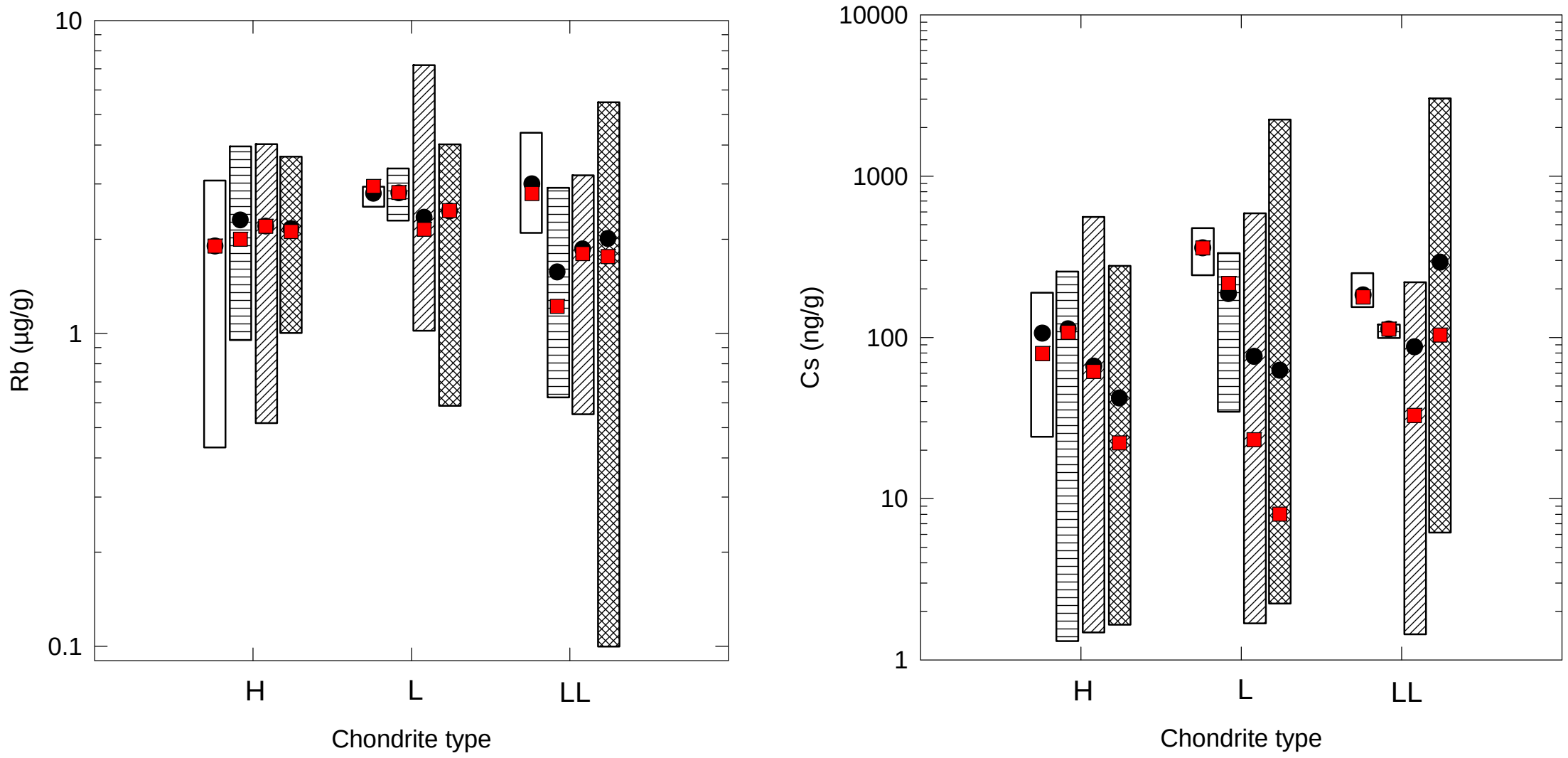

Figure 1. Rubidium and Cesium abundances.

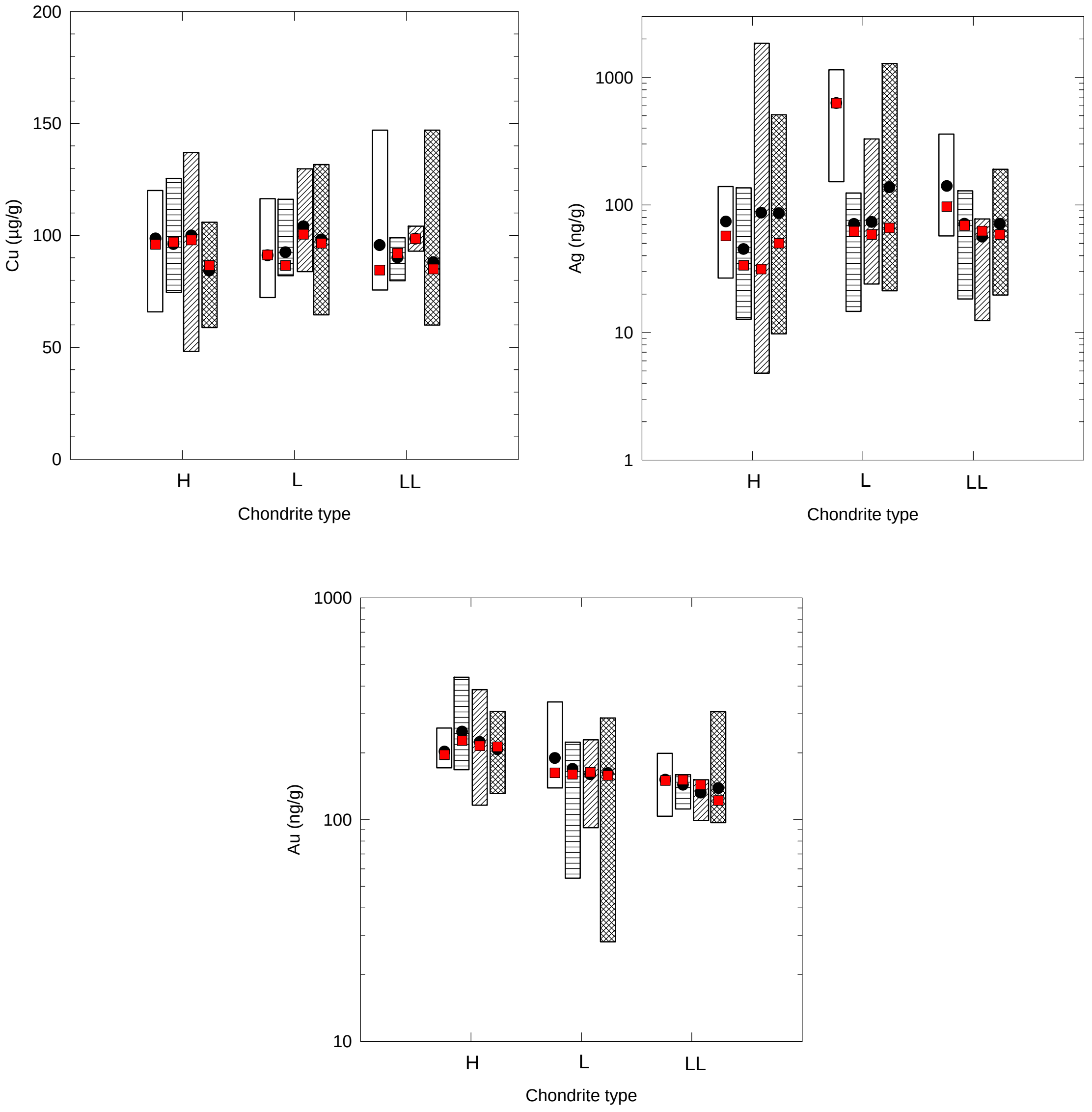

Figure 2: Abundances of Cu, Ag, and Au.

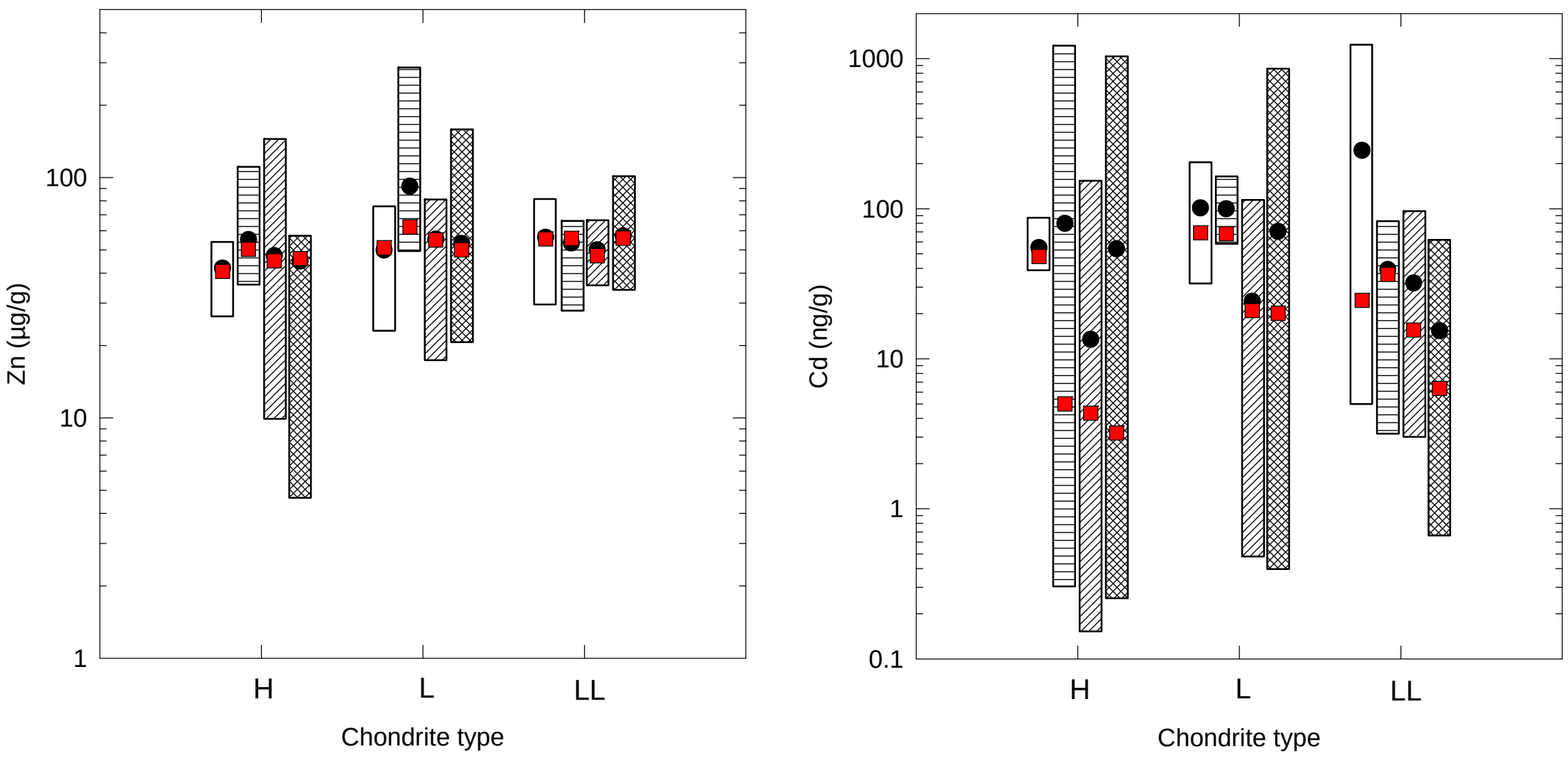

Figure 3: Abundances of Zn and Cd.

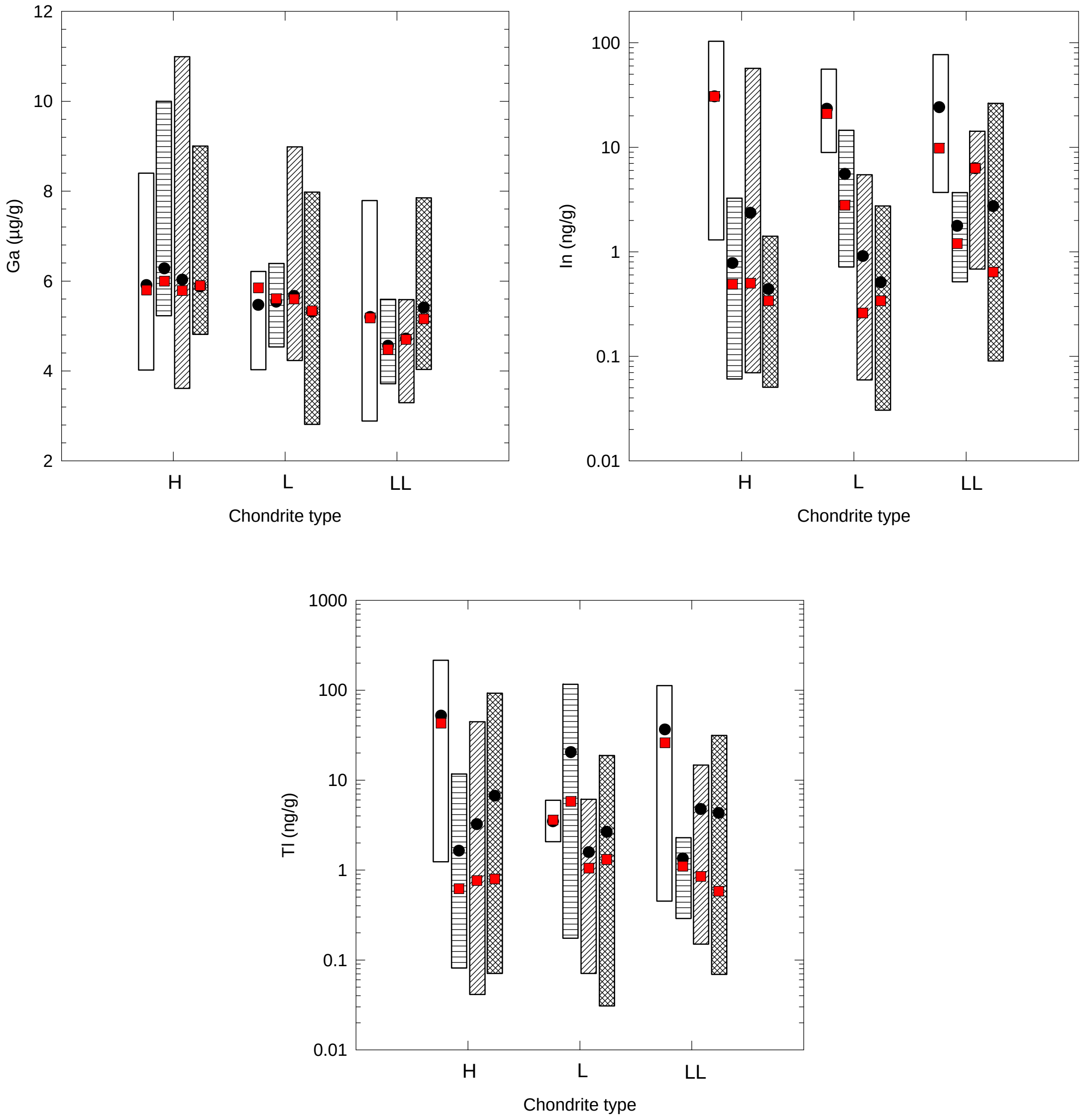

Figure 4: Abundances of Ga, In, and Tl.

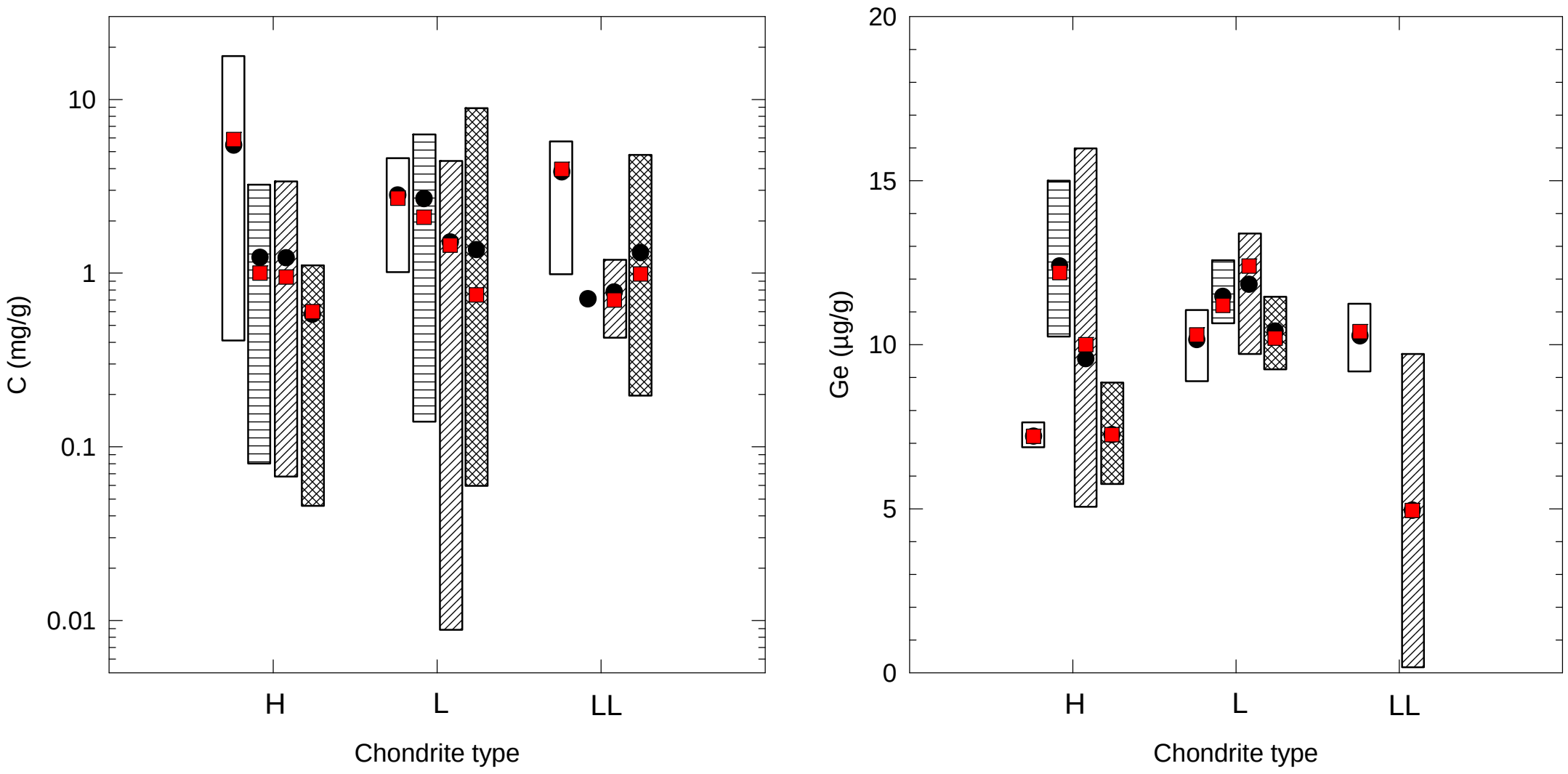

Figure 5: Abundance of (a) C, (b) Ge.

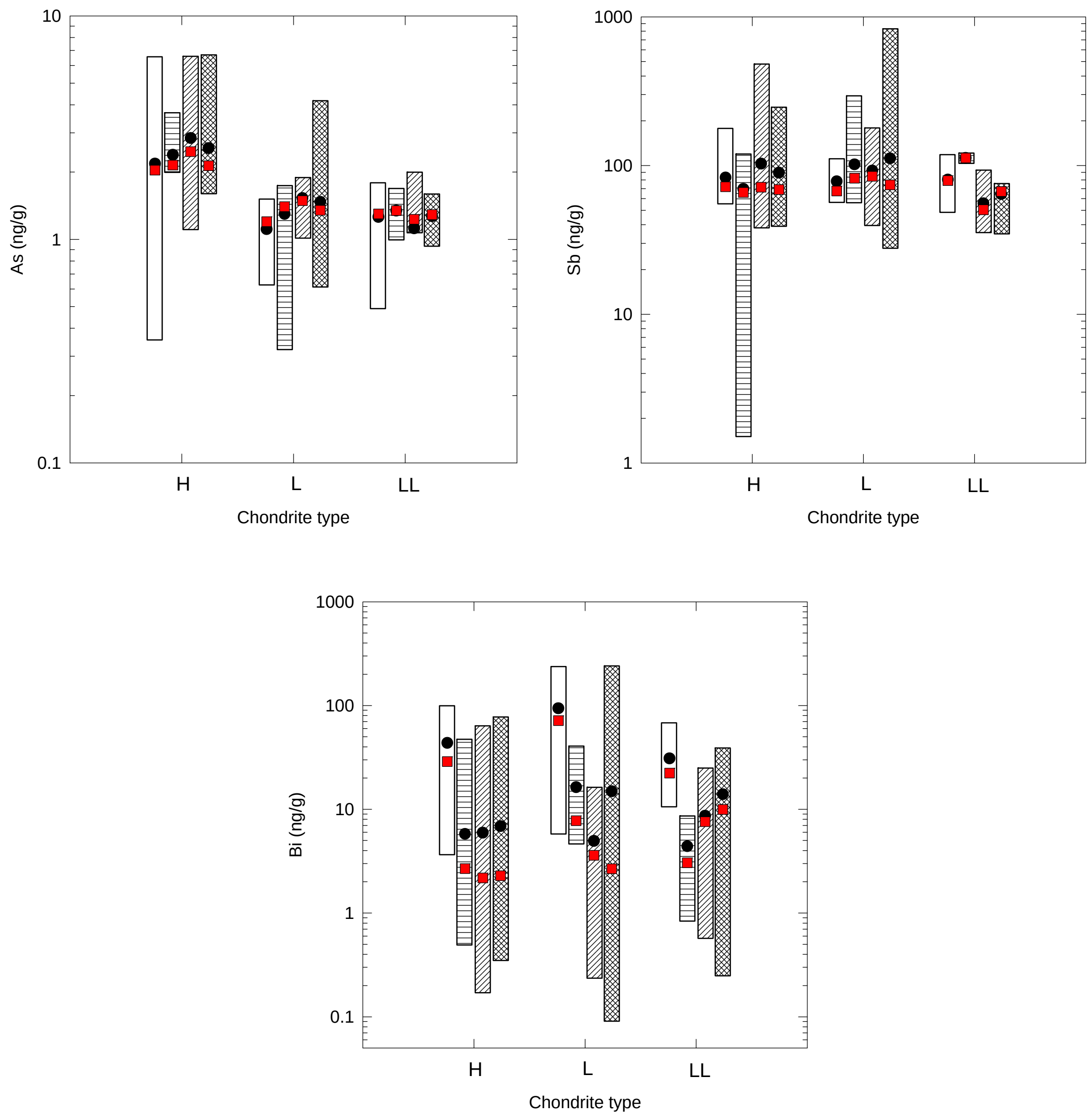

Figure 6: Abundances of As, Sb, and Bi.

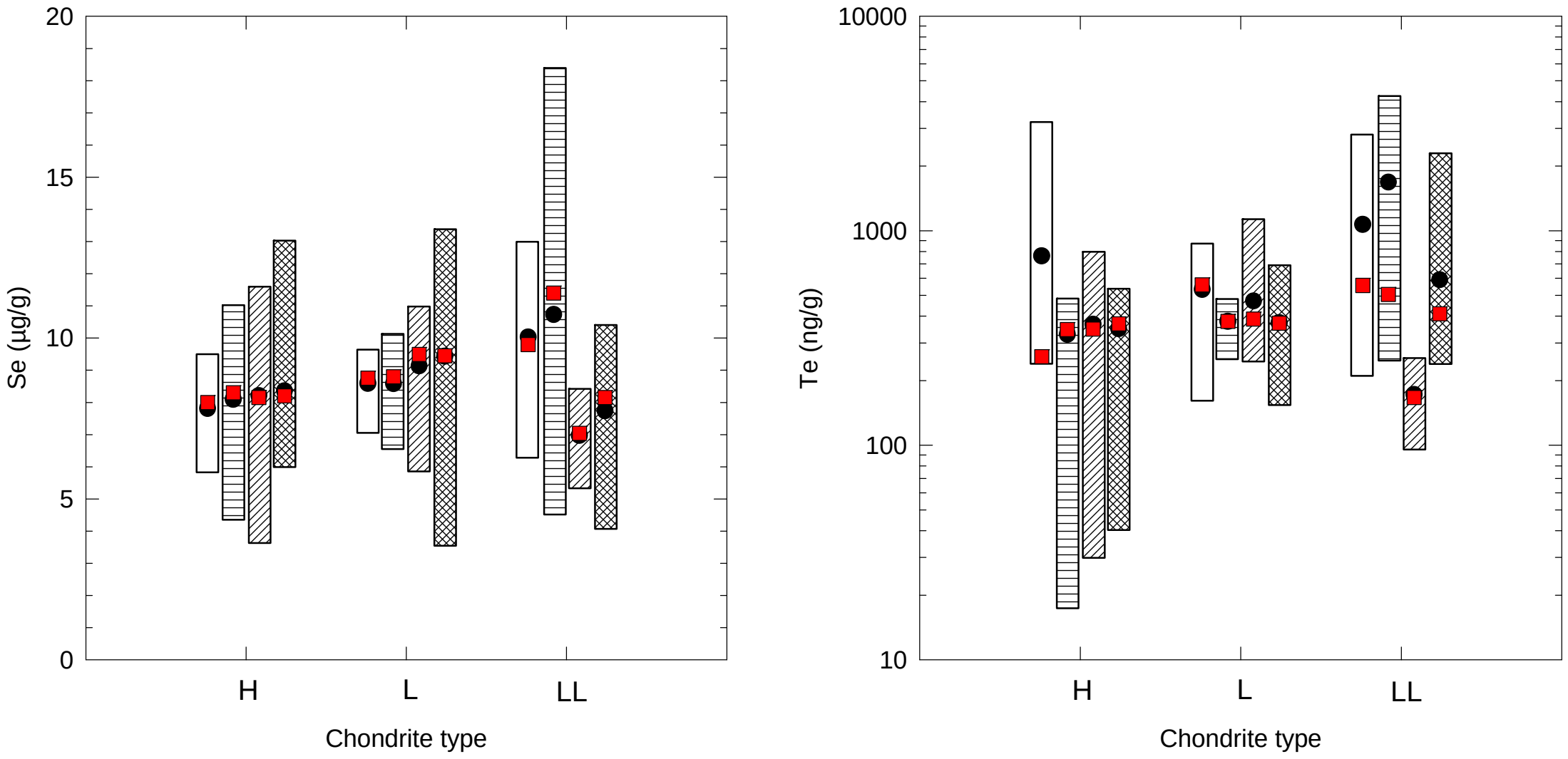

Figure 7: Abundances of Se and Te.

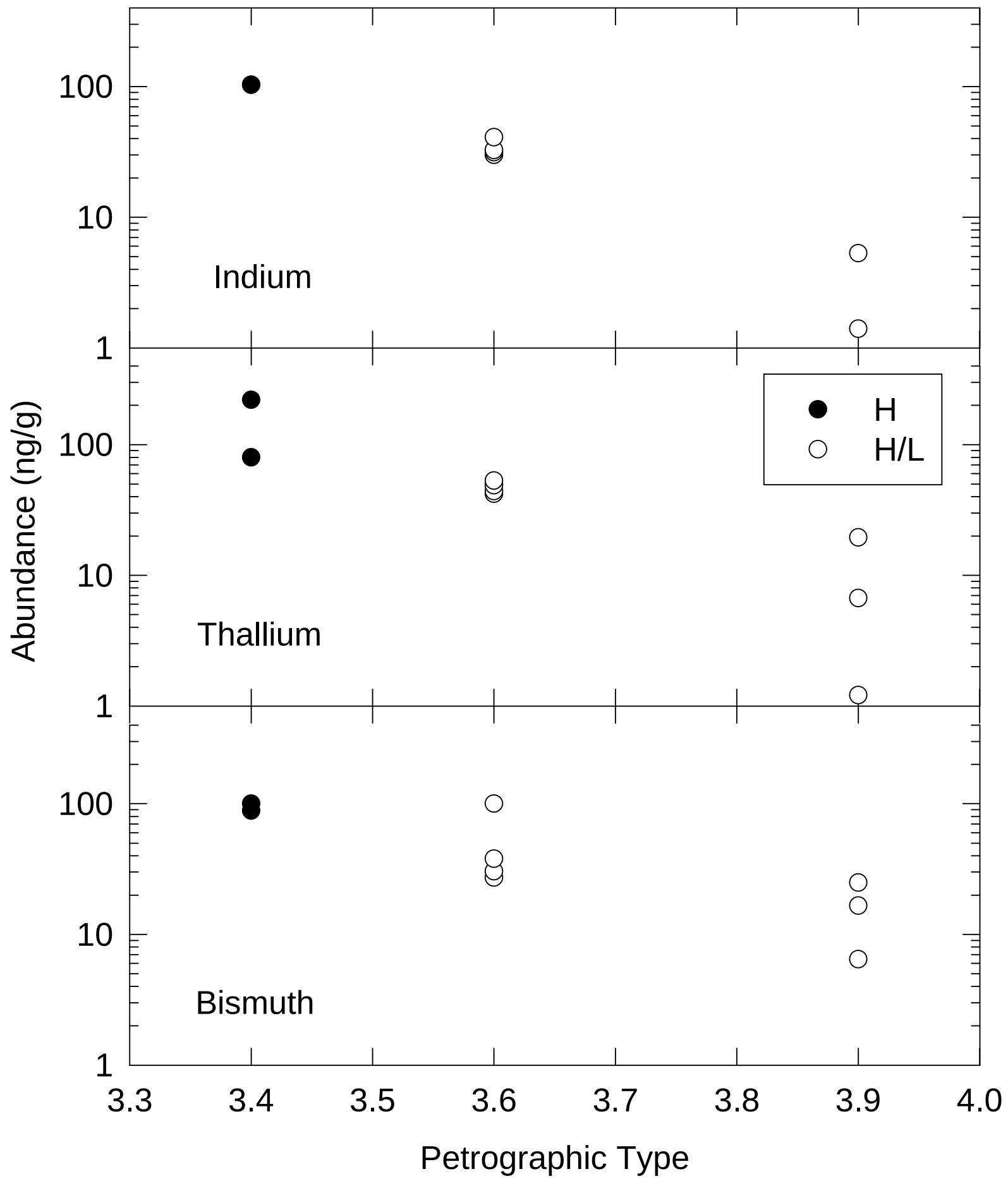

Figure 8 (a). In, Bi, Tl in type H3 and H/L3 chondrites.

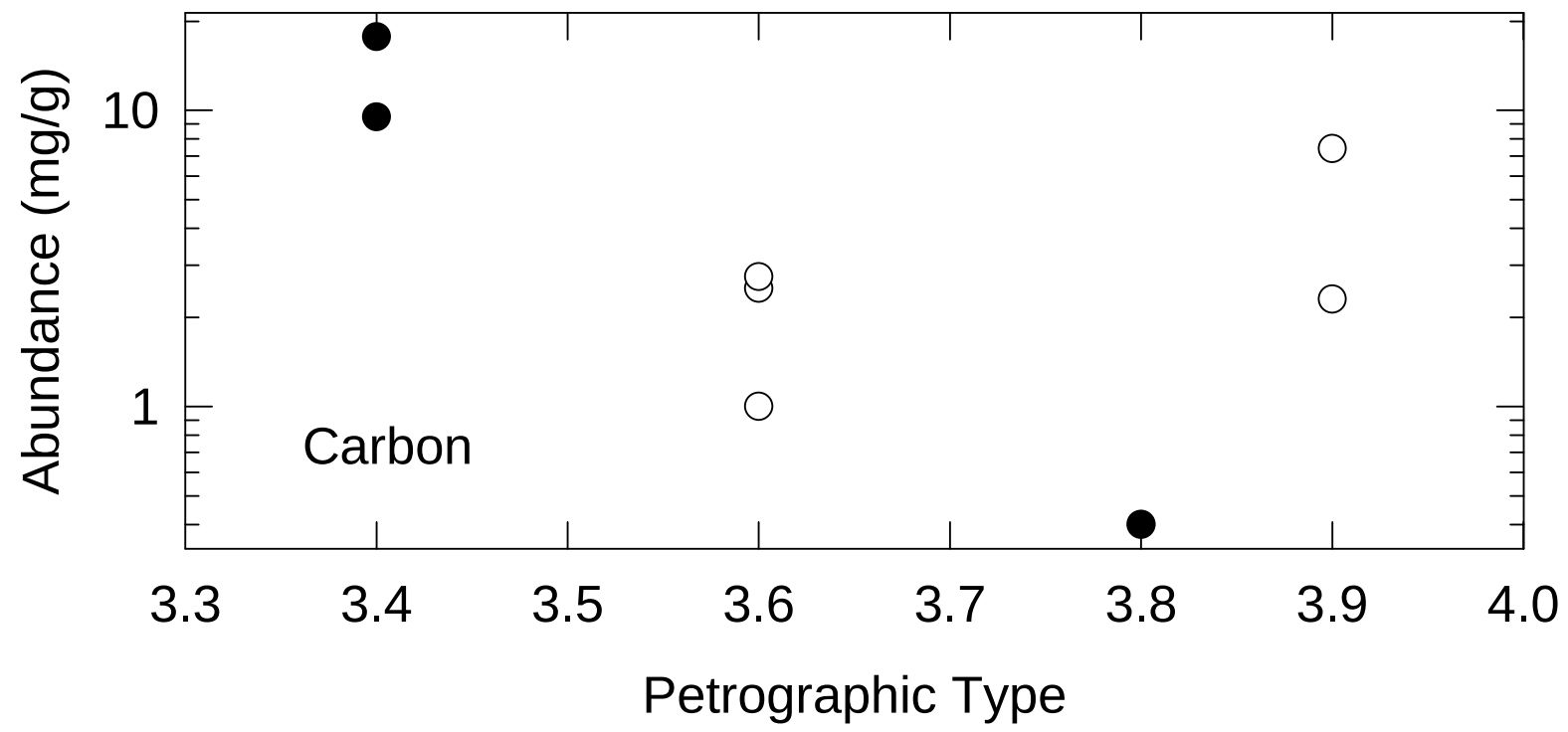

Figure 8 (b). C in type H3 and H/L3 chondrites.

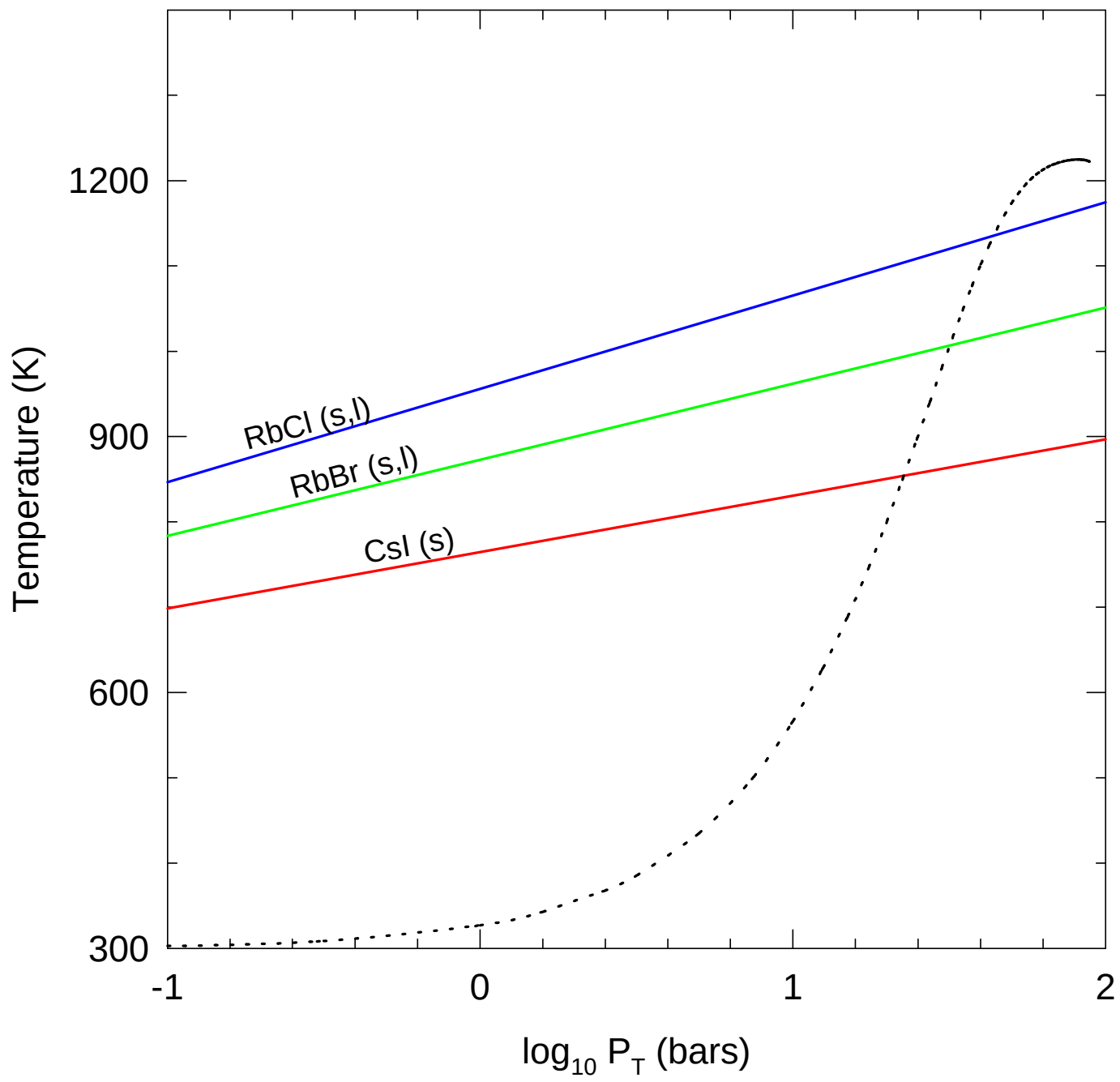

Figure 9. Rb, Cs chemistry in H-chondrites

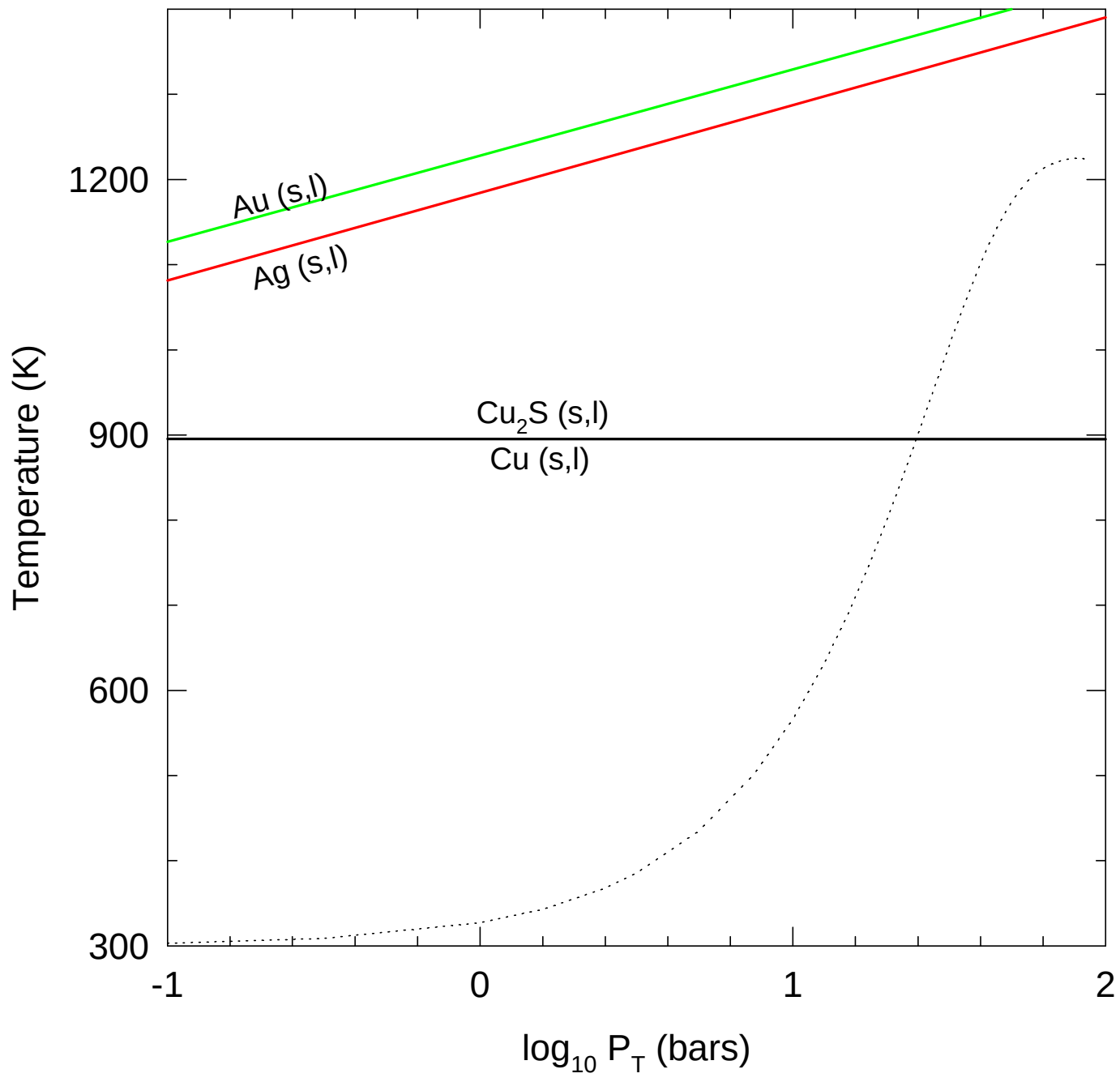

Figure 10. Cu, Ag, Au chemistry in H-chondrites

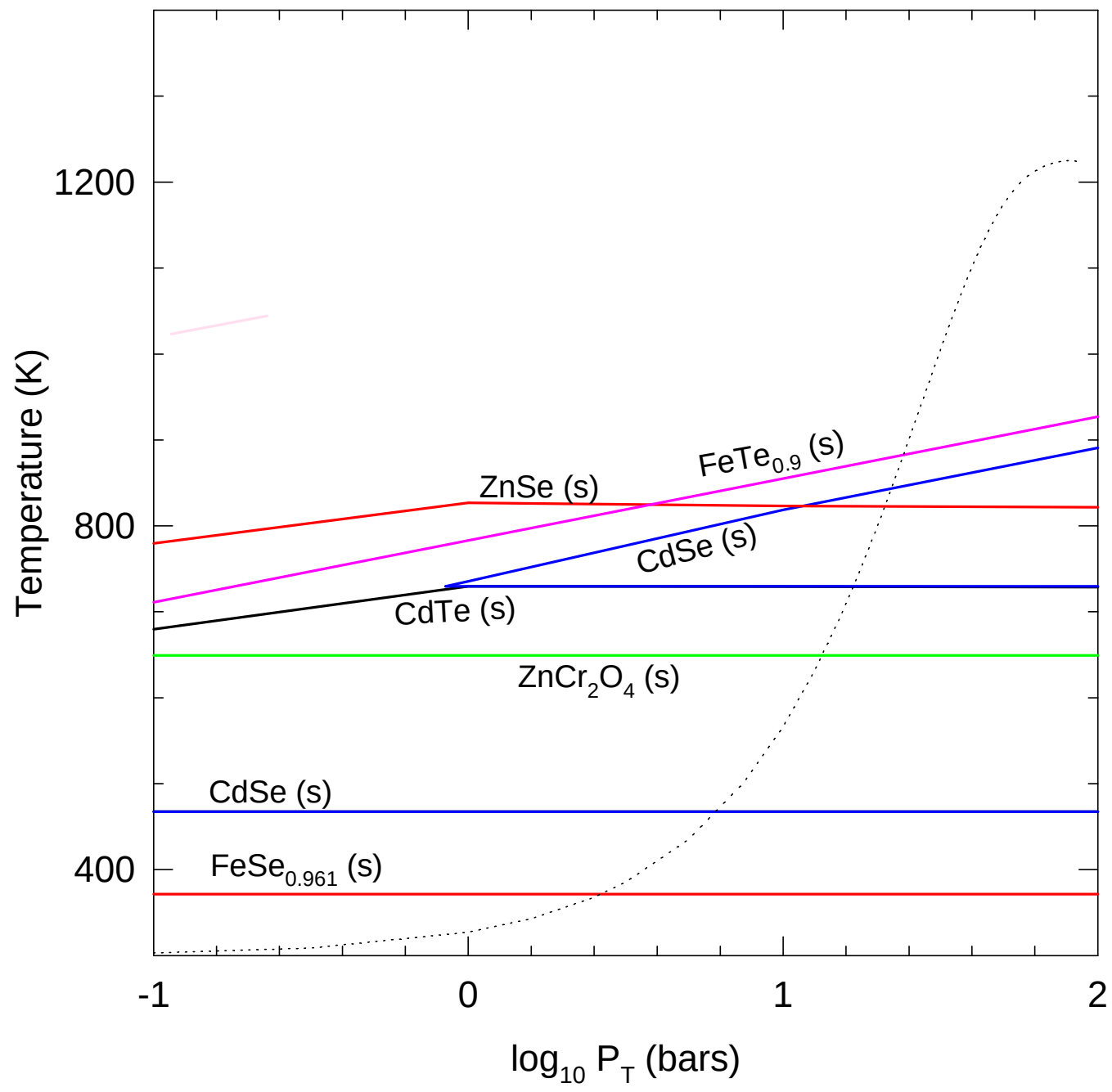

Figure 11. Group IIB: Zn, Cd and Group VIA: Se, Te.

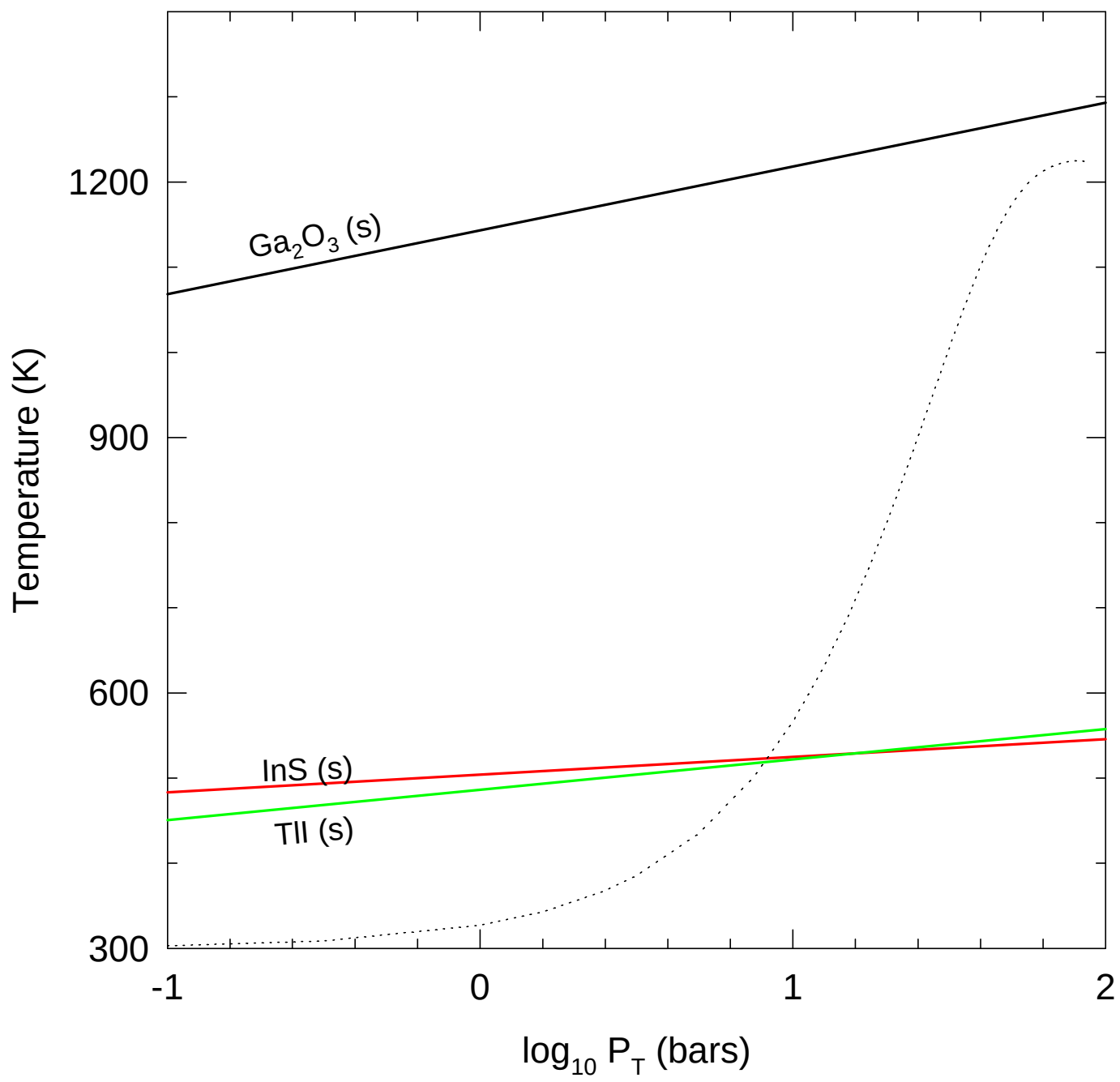

Figure 12. Group IIIA: Ga, In, Tl

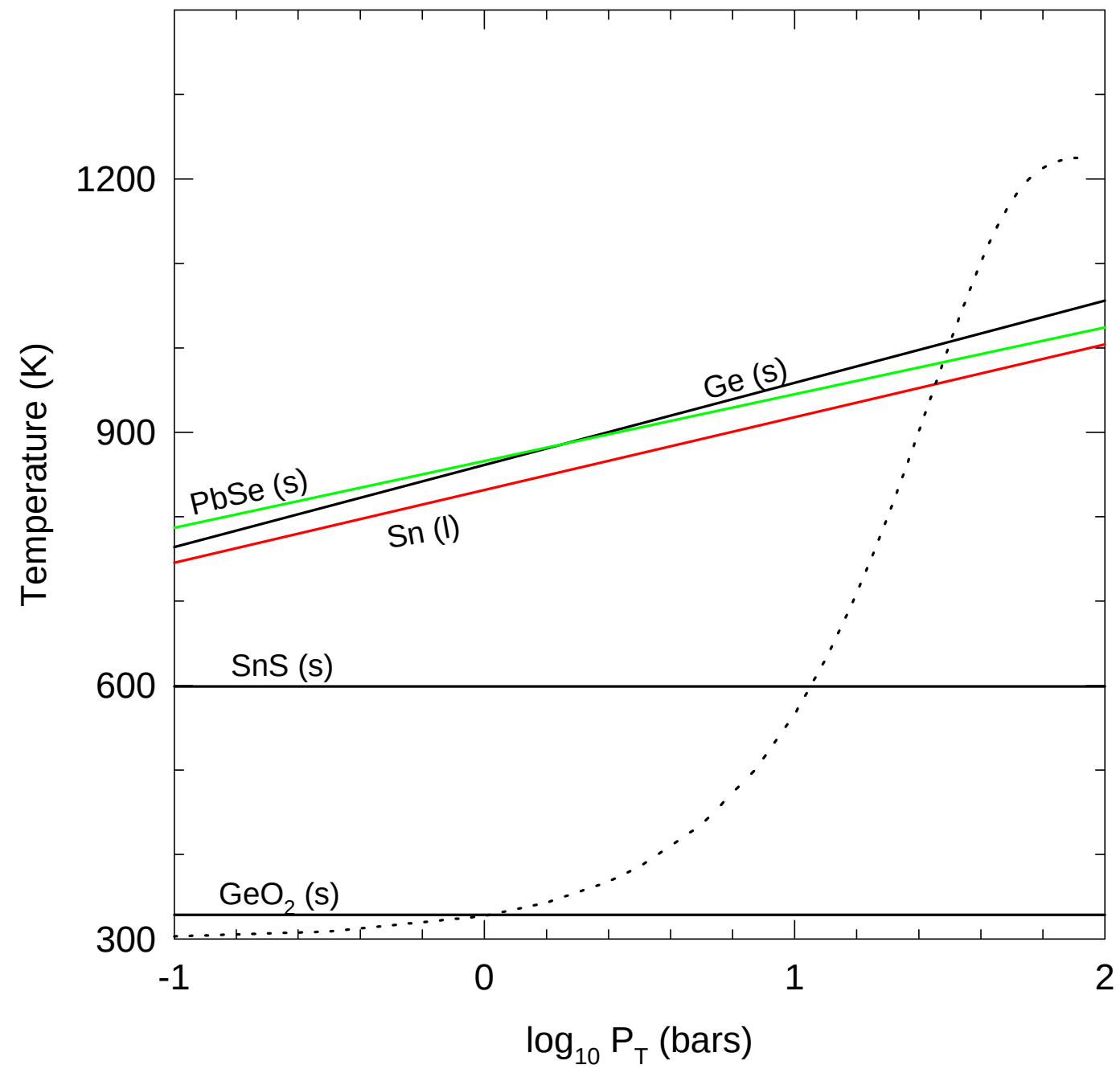

Figure 13. Group IVA: Ge, Sn, Pb

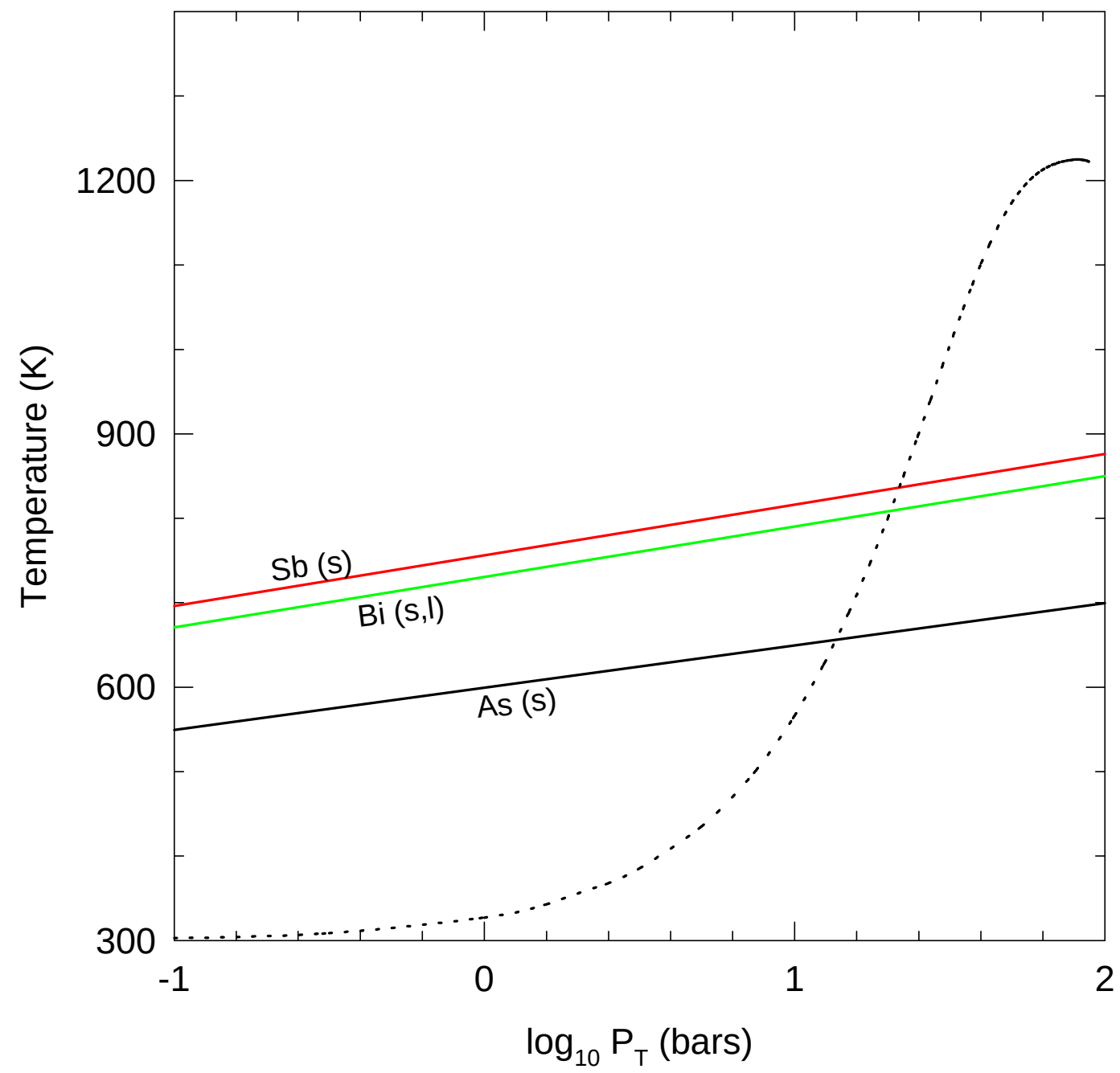

Figure 14. Group VA: As, Sb, Bi